\documentclass[aps,pre,reprint,longbibliography,superscriptaddress]{revtex4-2}

\usepackage[T1]{fontenc}
\usepackage[utf8]{inputenc}
\usepackage{amsmath,amssymb,amsfonts,bm,mathtools}
\usepackage{physics}
\usepackage{graphicx}
\usepackage{xcolor}
\usepackage{booktabs}
\usepackage{multirow}
\usepackage{hyperref}
\usepackage{enumitem}
\usepackage{comment}
\usepackage{ragged2e}

\usepackage{orcidlink}

\usepackage{tikz}

\hypersetup{colorlinks=true,linkcolor=blue,citecolor=blue,urlcolor=blue}

\newcommand{\Var}{\operatorname{Var}}

\begin{document}
\title{Coherence-Preserving Fluctuation Diagnostics for an Engineered Population-Inverted Qubit Otto Engine}

\author{Gabriella G. Damas~\orcidlink{0000-0003-3376-9281}}
\affiliation{Department of Physics, Zhejiang Normal University, Jinhua 321004, China}
\affiliation{Instituto de Física, Universidade Federal de Goiás, 74.001-970, Goiânia
- GO, Brazil}

\author{Norton G. de Almeida~\orcidlink{0000-0001-8517-6774}}%\email{norton@ufg.br}
\affiliation{Instituto de Física, Universidade Federal de Goiás, 74.001-970, Goiânia
- GO, Brazil}

\author{Gao Xianlong~\orcidlink{0000-0001-6914-3163}}
 \email{gaoxl@zjnu.edu.cn}
\affiliation{Department of Physics, Zhejiang Normal University, Jinhua 321004, China}

\author{G. D. de Moraes Neto~\orcidlink{0000-0003-4273-8380}}
 \email{gdmneto@gmail.com}
\affiliation{Faculty of Civil Engineering and Mechanics, Kunming University of Science and Technology, Kunming, 650500, China}

\date{\today}

\begin{abstract}
Finite‑time quantum thermal machines require diagnostics beyond average work and efficiency, 
because microscopic engines operate in regimes where fluctuations, incomplete thermalization, and coherence are equally important.  
Here we develop a measurement‑backaction‑free (coherence‑preserving) fluctuation diagnostic for an engineered qubit Otto engine 
coupled to an actively maintained population‑inverted hot channel.  
The engine is analyzed using a dynamic Bayesian network (DBN) reconstruction of the unmeasured coherent cycle, 
yielding work, heat, power, and normalized efficiency‑proxy fluctuations without imposing the projective dephasing inherent in two‑point energy measurements.  
The inverted channel is treated as an active reduced‑model resource; accordingly, all reported power and efficiency enhancements represent gross working‑medium advantages, not net device efficiencies.  
In the full‑thermalization limit, population inversion enhances extracted work and output power while opening a stability sector with markedly reduced relative power fluctuations.  
When finite‑duration hot and cold isochores are implemented, this gross enhancement reorganizes into a structured operating landscape with distinct high‑power, high‑efficiency, and low‑relative‑noise sectors, whose boundaries are understood through the competing timescales of nonadiabatic driving and thermalization rates. 
A direct comparison with the conventional two‑point measurement scheme reveals that DBN and TPM predictions diverge precisely in coherence‑rich regimes, 
identifying the operating sectors where a backaction‑free reconstruction is essential.  
A coherence‑sensitive analysis further shows that the positive‑temperature reference operates optimally in an almost decohered region, whereas the inverted high‑efficiency branch remains aligned with the dominant post‑hot‑bath coherence ridge.  
These results provide a reduced‑model benchmarking framework for engineered qubit thermal machines, pinpointing regimes where active reservoir engineering, 
fluctuation stability, and coherence‑preserving reconstruction become simultaneously relevant.
\end{abstract}

\maketitle

\section{Introduction}

Quantum thermodynamics provides the conceptual framework for describing heat, work, and irreversibility in microscopic devices, where quantum coherence, engineered dissipation, and nonequilibrium reservoirs can modify the operation of thermal machines beyond classical expectations \cite{Alicki2015}. Among the available models, the quantum Otto engine has become a central platform because it combines conceptual simplicity with experimental relevance: it consists of two unitary strokes and two isochoric contacts, and has been studied in qubit, spin, optical, and engineered-reservoir settings \cite{Campisi2011, Roulet2018, deAssis2019, Matos2023, XuHe2024, Xu2024, Zhou2026,N1,N2,N3}. For quantum-device applications, however, average work and efficiency are not sufficient performance indicators. Finite-time operation, nonadiabatic driving, and microscopic noise make work, heat, power, and efficiency fluctuations intrinsic diagnostics of stability and reliability \cite{Manikandan2019,Denzler2020, de2021two,Denzler2024,Anka2024,Gramajo2023}. Recent experiments have further shown that work--heat correlations, efficiency statistics, and entropy-production fluctuations can now be accessed in controlled quantum platforms \cite{Batalhao2014,Solfanelli2021,Denzler2024}.

A central difficulty is that work is not represented by a standard quantum observable, so its statistics depend on the operational reconstruction protocol \cite{Talkner2007,Esposito2009,Talkner2016}. The two-point measurement (TPM) scheme provides a consistent description of explicitly measured energy changes and underlies much of the fluctuation-theorem literature \cite{Talkner2007,Esposito2009,Talkner2016,Batalhao2014,Solfanelli2021}. However, for coherent finite-time engines, the initial projective energy measurement dephases the working medium and can change the subsequent dynamics \cite{Perarnau2017,Xu2018,Micadei2020,Lostaglio2018,Landi2024}. This is especially relevant when nonadiabatic strokes generate coherence that contributes to inner friction, work extraction, or performance losses \cite{Camati2019,Brandner2017,Santos2019,Francica2019}. Several alternatives to TPM have therefore been developed, including interferometric and ancilla-assisted reconstructions, work meters, weak and continuous measurements, full counting statistics, quasiprobabilities, and histories-based approaches \cite{Dorner2013,Mazzola2013,Hayashi2017,Batalhao2014,Cerisola2017,Prasanna2015,Elouard2017,Hekking2013,Landi2024,Nazarov2003,Allahverdyan2014,Solinas2015,Solinas2016,Hofer2016,Hofer2017,Miller2017,Lostaglio2023,Gherardini2021,Gherardini2024}. These developments make clear that different prescriptions need not agree in the presence of coherence, and that no universal measurement scheme can simultaneously preserve the untouched coherent average work and reproduce TPM statistics for all incoherent states \cite{Allahverdyan2014,Solinas2016,Hofer2016,Hofer2017,Xu2018,Lostaglio2018,Lostaglio2023,Gherardini2024,Perarnau2017}.

In this work we adopt the dynamic Bayesian network (DBN) framework to reconstruct fluctuation statistics associated with the unmeasured coherent dynamics of a finite-time qubit Otto engine. DBN uses projectors onto the eigenbases of the corner density operators and reconstructs multi-time path probabilities while preserving the average state and average energetics of the untouched cycle \cite{Micadei2020,Micadei2021,Micadei2023,Strasberg2019,Strasberg2020,Park2020,Zhang2022,Rodrigues2024,Aguilar2026}. This choice should not be read as a universal replacement for TPM. Rather, TPM describes the explicitly measured and dephased machine, whereas DBN is an ensemble-level reconstruction protocol for the coherent, effectively unmeasured cycle. This distinction is essential because coherence can either enhance or suppress thermodynamic performance depending on how it is generated, processed, and dissipated during the cycle \cite{Scully2011,Brandner2017,Camati2019,Santos2019,Francica2019,Rodrigues2024}.

The second ingredient of the present study is an actively maintained population-inverted hot channel. Negative temperatures were historically introduced for bounded spectra \cite{Purcell1951,Ramsey1956}, and modern treatments describe population-inverted states as effective or apparent negative-temperature resources \cite{Struchtrup2018,Bera2024}. Such resources have been considered in autonomous refrigerators and heat engines \cite{Damas2023,Xi2017,deAssis2019,Nettersheim2022,desousa2024}. Nevertheless, an inverted reservoir is not a passive heat bath: maintaining it generally requires pumping, active preparation, or a synthetic reservoir construction. The results below should therefore be interpreted as gross working-medium performance within a reduced inverted-channel model, not as a complete resource-accounted device efficiency.

The central question addressed here is how active population inversion, finite-time thermalization, and coherence-preserving fluctuation reconstruction jointly shape the operating landscape of an engineered qubit Otto engine. In the full-thermalization benchmark, the inverted channel isolates the role of coherent driving and population inversion. In the finite-time regime, incomplete hot and cold isochores allow memory and coherence to survive across strokes, making the engine genuinely history dependent. We therefore analyze mean extracted work and power together with work fluctuations, relative power fluctuations, and a normalized efficiency proxy. We also use DBN-based coherence diagnostics and recently proposed information-theoretic and fluctuation-based bounds as consistency checks \cite{Rignon-Bret2021,Souza2022,Saryal2021,Rodrigues2024,Aguilar2026}.

From a technology-oriented perspective, the purpose of the paper is to provide a reduced-model diagnostic framework rather than a complete heat-engine device proposal. The finite-time maps identify operating sectors with high gross power, sectors with lower relative power noise, and sectors where coherence-preserving reconstruction is most relevant. Section~II introduces the model and thermodynamic observables. Section~III presents the DBN construction and fluctuation diagnostics. Section~IV analyzes the full-thermalization benchmark, and Sec.~V studies the finite-time operating landscapes. We conclude by discussing the physical interpretation, limitations, and possible platform-specific extensions.

\section{Model and observables}
\label{sec:model}

\subsection{Working medium and Otto cycle}
\label{sec:model_cycle}

\begin{figure}[t!] 
    \centering
    \includegraphics[width=0.47\textwidth]{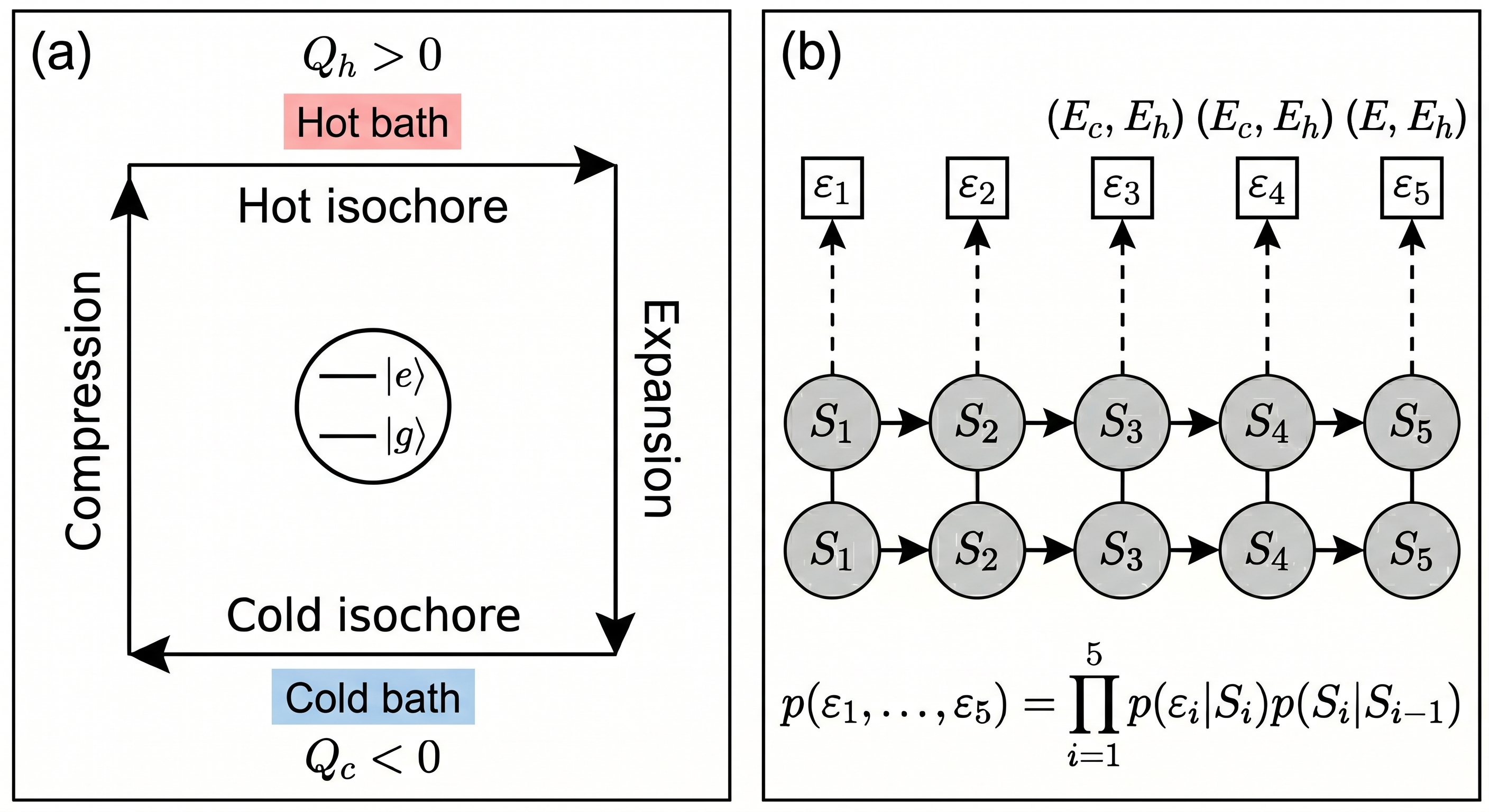}
 \caption{\justifying (a) Schematic of the finite-time qubit Otto cycle considered in this work. The cycle consists of a compression stroke, a hot isochore, an expansion stroke, and a cold isochore. During the hot isochore the engine absorbs heat from the hot reservoir, $Q_h>0$, while during the cold isochore it releases heat to the cold reservoir, $Q_c<0$. (b) Dynamic Bayesian network representation of the five-corner cycle. The lower chain represents the sequence of corner-state variables along the cycle, while the upper nodes represent the inferred corner-energy outcomes \(\epsilon_i\). The horizontal arrows describe the temporal propagation of the corner-state variables through the cycle, whereas the vertical dashed arrows indicate that the energy variables are inferred conditionally from the corresponding corner states rather than obtained from direct projective measurements. This is the probabilistic structure underlying the DBN path distribution used in the fluctuation analysis.
}
    \label{fig:otto_dbn}
    \end{figure}

We consider a finite-time quantum Otto engine whose working medium is a two-level system. As illustrated in Fig.~\ref{fig:otto_dbn}(a), the cycle consists of two unitary strokes, compression and expansion, and two isochoric strokes, a hot isochore and a cold isochore. The cold and hot Hamiltonians are
\begin{equation}
H_c=-\frac{\omega_c}{2}\sigma_x,
\qquad
H_h=-\frac{\omega_h}{2}\sigma_y,
\qquad
\omega_h>\omega_c>0,
\label{eq:Hch}
\end{equation}
so that the unitary branches connect two noncommuting energy bases and may dynamically generate coherence.

The compression stroke drives the system from $H_c$ to $H_h$ during a time $\tau_{\rm drive}$. The hot isochore keeps the Hamiltonian fixed at $H_h$ while the qubit interacts with the hot reservoir during a time $\tau_h$. The expansion stroke drives the system back from $H_h$ to $H_c$ during a second interval of duration $\tau_{\rm drive}$. Finally, during the cold isochore, the Hamiltonian is fixed at $H_c$ and the qubit interacts with the cold reservoir during a time $\tau_c$. The total cycle time is therefore
\begin{equation}
\tau_{\rm cyc}=2\tau_{\rm drive}+\tau_h+\tau_c.
\label{eq:taucyc}
\end{equation}
For the unitary branches we use a smooth finite-time interpolation in the $(\sigma_x,\sigma_y)$ plane, in close analogy with the driven qubit Otto protocol of Ref.~\cite{deAssis2019}. Writing
\begin{equation}
\theta(t)=\frac{\pi t}{2\tau_{\rm drive}},
\end{equation}
the compression and expansion Hamiltonians are taken as
\begin{align}
H_{\rm comp}(t)
&=
-\frac{\Omega_{c\to h}(t)}{2}
\left[
\cos\theta(t)\,\sigma_x+\sin\theta(t)\,\sigma_y
\right],
\\
H_{\rm exp}(t)
&=
-\frac{\Omega_{h\to c}(t)}{2}
\left[
\sin\theta(t)\,\sigma_x+\cos\theta(t)\,\sigma_y
\right],
\end{align}
with linearly varying gaps
\begin{align}
\Omega_{c\to h}(t)
&=
\omega_c\!\left(1-\frac{t}{\tau_{\rm drive}}\right)
+\omega_h\frac{t}{\tau_{\rm drive}},
\\
\Omega_{h\to c}(t)
&=
\omega_h\!\left(1-\frac{t}{\tau_{\rm drive}}\right)
+\omega_c\frac{t}{\tau_{\rm drive}}.
\end{align}
The corresponding propagators are denoted by $U_{c\to h}$ and $U_{h\to c}$.
\subsection{Reservoir models and effective negative temperature}
\label{sec:model_baths}

During each isochore the state evolves according to a Markovian master equation at fixed Hamiltonian,
\begin{equation}
\dot{\rho}
=
-i[H_\alpha,\rho]
+\sum_k \mathcal{D}[L_k]\rho,
\qquad
\alpha\in\{c,h\},
\label{eq:ME}
\end{equation}
where $\mathcal{D}[L]\rho=L\rho L^\dagger-\frac{1}{2}\{L^\dagger L,\rho\}$. The collapse operators are defined in the instantaneous energy basis of $H_c$ or $H_h$.

For a positive-temperature bosonic implementation, the steady excited-state population can be written in terms of the Bose occupation number
\begin{equation}
\bar n_\alpha=\frac{1}{e^{\beta_\alpha\omega_\alpha}-1},
\qquad
p_{e,\alpha}^{\rm(B)}=
\frac{\bar n_\alpha}{2\bar n_\alpha+1},
\label{eq:bose_pop}
\end{equation}
where $\omega_\alpha\in\{\omega_c,\omega_h\}$. This expression satisfies $0\le p_{e,\alpha}^{\rm(B)}<1/2$ for a passive positive-temperature bosonic bath. Thus the inverted regime studied below should not be read as a standard bosonic thermal reservoir. A bounded spin reservoir, a synthetic reservoir, an actively pumped two-level environment, or a fermionic/spin-temperature channel is required to realize $p_e>1/2$.

For a fermionic or two-level bath with detailed-balance form,
\begin{equation}
f_\alpha=\frac{1}{e^{\beta_\alpha\omega_\alpha}+1},
\qquad
p_{e,\alpha}^{\rm(F)}=f_\alpha,
\label{eq:fermi_pop}
\end{equation}
negative inverse temperature corresponds to $p_{e,\alpha}^{\rm(F)}>1/2$. Equivalently, the Markovian dissipator may be parametrized by upward and downward rates
\begin{equation}
	\begin{aligned}
		\Gamma_{\uparrow,\alpha} &= \gamma_\alpha p_{e,\alpha}, \\
		\Gamma_{\downarrow,\alpha} &= \gamma_\alpha(1-p_{e,\alpha}), \\
		\frac{\Gamma_{\uparrow,\alpha}}{\Gamma_{\downarrow,\alpha}}
		&=
		\frac{p_{e,\alpha}}{1-p_{e,\alpha}}
		=
		e^{-\beta_\alpha\omega_\alpha}.
	\end{aligned}
	\label{eq:rate_ratio}
\end{equation}
For $\gamma_\alpha>0$ and $0\leq p_{e,\alpha}\leq1$, both rates are non-negative, so the corresponding two-level amplitude-damping generator remains of GKSL form and therefore generates a completely positive trace-preserving Markovian evolution. The inverted regime changes the ratio of upward to downward rates, $\Gamma_{\uparrow,\alpha}/\Gamma_{\downarrow,\alpha}>1$, but does not by itself violate complete positivity. The physical nontriviality lies instead in the active reservoir engineering required to realize and maintain such rates.
For a bounded two-level spectrum, the condition
\begin{equation}
p_e^h>\frac{1}{2}
\label{eq:negtemp}
\end{equation}
therefore corresponds to $\beta_h<0$ and to population inversion \cite{Struchtrup2018,Bera2024}. Experimentally, the closest realizations of such an effective resource include spin-temperature reservoirs in NMR engines, quasi-spin engines coupled to engineered atomic baths, and synthetic reservoirs constructed with auxiliary driven degrees of freedom or multiple positive-temperature resources \cite{deAssis2019,Nettersheim2022,Bera2024}. In the present work we do not select one specific implementation. The inverted Lindblad channel should instead be read as the reduced working-medium description that would result after such reservoir engineering and active stabilization.

The finite-time scans use this inverted Markovian channel as an effective reduced description. The required inversion is substantial: the working point $p_e^h=0.80$ used in the finite-time maps corresponds to $\Gamma_\uparrow/\Gamma_\downarrow=4$, while the ideal-reset point $p_e^h=0.98$ corresponds to $\Gamma_\uparrow/\Gamma_\downarrow=49$. These numbers make explicit that the hot stroke is an active resource rather than an ordinary passive bath.

Throughout this work, the thermalization strokes are analyzed within a weak-coupling time-local Markovian description. At sufficiently short bath-contact times, non-Markovian memory, reservoir depletion, pump-induced heating, leakage, and saturation effects may become relevant and could quantitatively modify the shortest-time sector of the finite-time landscape. Because no explicit pump or reservoir-preparation model is included, all efficiencies and powers reported below are gross quantities for the reduced working-medium dynamics. A fully resource-accounted device efficiency would require adding the external cost of maintaining the inversion, for example through a pump work or exergy term, and is left outside the scope of the present reduced-model study.

\subsection{Quantum-technological interpretation and candidate platforms}

The population-inverted channel used in this work should be understood as an effective reservoir-engineering primitive rather than as an ordinary equilibrium heat bath. This interpretation is important for connecting the reduced model to possible quantum-device implementations. Several platforms could realize different parts of the required physics. Spin-temperature platforms, such as NMR systems, provide a natural setting for effective negative temperatures and full state reconstruction \cite{deAssis2019}. Cold-atom or trapped-atom quasi-spin engines offer controllable finite-time strokes and engineered dissipative contacts \cite{Nettersheim2022}. Superconducting circuits provide fast qubit control, tunable dissipation, and high-fidelity state tomography, making them a promising route for future implementations of active-reservoir qubit engines. Synthetic-reservoir constructions based on auxiliary quantum systems provide another possible route to effective population inversion \cite{Bera2024}.

For all these platforms, the present theory should be read as a diagnostic protocol rather than as a complete device proposal. An experimental DBN reconstruction would require repeated preparation of the stationary cycle, state reconstruction at the five corners $\rho_1,\ldots,\rho_5$, determination of the corresponding corner-state projector bases, and estimation of the conditional probabilities entering the Bayesian network. For a qubit, each corner state can in principle be reconstructed from three Pauli expectation values, but the procedure must be repeated, or at least recalibrated, for each operating point at which quantitative fluctuation statistics are desired. This makes the DBN protocol more demanding than a direct TPM experiment, but it targets a different object: the fluctuation statistics associated with the untouched coherent cycle rather than those of the explicitly measured and dephased cycle.

At the level of scaling, statistical uncertainty in single-qubit tomography decreases as $N_{\rm samp}^{-1/2}$ for $N_{\rm samp}$ repeated preparations per measurement setting. Since the DBN path probabilities are constructed from the reconstructed corner projectors and transition probabilities, these errors propagate into the inferred work and heat distributions. Quantifying this propagation for a specific platform is an important experimental step, but it is not included in the ideal reduced-model maps reported here.

The finite-time maps obtained below should therefore be interpreted as reduced-model design diagnostics. They identify which operating sectors are worth targeting in future platform-specific studies: high-gross-power sectors, low-relative-noise sectors, and coherence-aligned high-efficiency sectors. A complete device-level assessment would additionally require an explicit model of the pump or synthetic reservoir that maintains the inversion, together with sampling-error and tomography-noise propagation in the reconstructed DBN statistics.

\subsection{Mean thermodynamic observables and sign convention}
\label{sec:model_means}

We adopt the convention that work and heat are positive when added to the working medium. The mean first-law balance over one stationary cycle is therefore
\begin{equation}
\langle W\rangle+\langle Q_h\rangle+\langle Q_c\rangle=0.
\label{eq:firstlaw}
\end{equation}
In the engine regime,
\begin{equation}
\langle W\rangle<0,
\qquad
\langle Q_h\rangle>0,
\qquad
\langle Q_c\rangle<0,
\label{eq:engine_regime}
\end{equation}
so that negative work corresponds to work extraction.

The thermodynamic efficiency is defined as
\begin{equation}
\eta=-\frac{\langle W\rangle}{\langle Q_h\rangle},
\label{eq:eta}
\end{equation}
and the output power as
\begin{equation}
P=-\frac{\langle W\rangle}{\tau_{\rm cyc}}.
\label{eq:power}
\end{equation}
Because throughout this work we restrict attention to operating points satisfying $\langle Q_h\rangle>0$, the denominator in Eq.~\eqref{eq:eta} remains strictly positive in the engine regime. Full numerical parameters, scan grids, solver tolerances, and convergence checks are reported in Appendix \ref{app:numerics}.

\section{Dynamic Bayesian network description}
\label{sec:dbn}

The purpose of this section is to define the fluctuation reconstruction used to characterize the coherent finite-time Otto cycle. Since work is not represented by a standard quantum observable, different operational prescriptions correspond to different physical questions. The TPM scheme describes the statistics of explicitly measured energy changes. It is therefore the natural framework when projective energy measurements are part of the engine protocol. However, in a coherent finite-time engine, those measurements dephase the working medium in the instantaneous energy basis and can modify the subsequent dynamics. The resulting TPM statistics then refer to a measured and dephased machine, rather than to the untouched coherent cycle.

Here we instead use the DBN framework as an ensemble-level reconstruction of the unmeasured coherent dynamics. The DBN construction uses projectors onto the eigenbases of the corner density operators and infers the corresponding energy histories through conditional probabilities \cite{Micadei2020,Micadei2021,Rodrigues2024,Aguilar2026}. In this sense, the DBN path distribution is not a single-shot record of projective energy outcomes. Rather, it is a reconstructed multi-time distribution associated with repeated preparations of the stationary cycle. Its key role in the present work is that it preserves the average state and average energetics of the untouched dynamics, allowing fluctuations to be analyzed without imposing the dephasing associated with TPM.

This operational distinction is central to the interpretation of the results. DBN is not claimed to be a universal replacement for TPM, nor a backaction-free single-run measurement of work and heat. It answers a different question: what fluctuation statistics are inferred for the coherent cycle when the engine dynamics itself is not interrupted by projective energy measurements? Conversely, TPM remains the correct benchmark for the explicitly measured engine. The comparison between DBN and TPM should therefore be understood as a comparison between two physically distinct protocols, not as a purely technical difference between two estimators.

This choice also differs from quasiprobability and full-counting-statistics approaches \cite{Nazarov2003,Hekking2013,Solinas2015,Hofer2016,Hofer2017,Lostaglio2023,Gherardini2024}. Those formulations often preserve linearity in the quantum state and can encode coherence through negative or contextual probability assignments, which is advantageous for fluctuation-theorem analyses. The DBN construction used here instead keeps a positive Bayesian path distribution built from state-projector histories and conditional energy inference. Thus the price of DBN is the loss of a strictly classical trajectory-level first law, while its advantage for the present purpose is a positive ensemble-level reconstruction that preserves the untouched coherent averages of the cycle. We emphasize that \emph{coherence-preserving} in this context refers exclusively to avoiding the artificial measurement-induced dephasing injected by TPM at the stroke boundaries. The isochoric bath interactions themselves still naturally dissipate coherence according to the underlying completely positive and trace-preserving (CPTP) Markovian dynamics; DBN simply reconstructs the statistics of that physical dissipation without artificially interrupting it.

From an experimental perspective, DBN requires repeated preparation of the stationary cycle and prior reconstruction of the corner states. The local projectors entering the Bayesian network are determined by the density operators at the five cycle corners, so the reconstruction must be recalibrated when the operating point is changed. For a qubit, each corner state can in principle be reconstructed from Pauli tomography, but finite sampling and tomography errors would broaden the inferred distributions. These experimental overheads are not simulated here. The DBN results should therefore be read as ideal ensemble-level fluctuation diagnostics for the coherent reduced model, identifying the information that would be inaccessible or altered under direct projective energy measurements.

\subsection{DBN construction for the finite-time cycle}
\label{sec:dbn_cycle}

We describe one stationary cycle through the sequence of corner states
\begin{equation}
\rho_1 \rightarrow \rho_2 \rightarrow \rho_3 \rightarrow \rho_4 \rightarrow \rho_5,
\end{equation}
where $\rho_1$ is the state at the beginning of compression, $\rho_2$ the state after compression, $\rho_3$ the state after the hot isochore, $\rho_4$ the state after expansion, and $\rho_5$ the state after the cold isochore. The logic of the DBN construction is illustrated in Fig.~\ref{fig:otto_dbn}(b). The lower layer encodes the sequence of corner-state variables along the cycle, while the upper layer contains the inferred energy variables associated with each corner. The horizontal arrows describe the temporal propagation of the state variables through the cycle, whereas the vertical dashed arrows indicate that the energies are inferred conditionally from the corresponding corner states. In the notation of the figure, the lower nodes $S_i$ schematically represent the corner-state variables associated with the state-projector outcomes $\alpha_i$, while the upper nodes $\epsilon_i$ denote the inferred corner-energy variables.

For the finite-time engine,
\begin{align}
\rho_2 &= U_{c\to h}\rho_1 U_{c\to h}^\dagger,
\\
\rho_3 &= \mathcal{E}_h(\rho_2),
\\
\rho_4 &= U_{h\to c}\rho_3 U_{h\to c}^\dagger,
\\
\rho_5 &= \mathcal{E}_c(\rho_4),
\end{align}
with $\rho_1$ chosen as the fixed point of the full cycle map,
\begin{equation}
\Lambda
=
\mathcal{E}_c \circ \mathcal{U}_{h\to c} \circ \mathcal{E}_h \circ \mathcal{U}_{c\to h},
\qquad
\Lambda(\rho_1)=\rho_1.
\label{eq:cycle_fixed_point}
\end{equation}

At each corner $a=1,\dots,5$, we consider both the spectral decomposition of the state,
\begin{equation}
\rho_a=\sum_{\alpha_a}\lambda_{\alpha_a}^{(a)}P_{\alpha_a}^{(a)},
\label{eq:rho_corner_spec}
\end{equation}
and the spectral decomposition of the corresponding Hamiltonian,
\begin{equation}
H_a=\sum_{j_a}e_{j_a}^{(a)}\Pi_{j_a}^{(a)}.
\label{eq:H_corner_spec}
\end{equation}
In the DBN language, the state-projector outcomes define the underlying chain of corner-state variables, while the energy variables are inferred conditionally from them, as indicated schematically in Fig.~\ref{fig:otto_dbn}(b). The conditional probabilities for the corner energies are
\begin{equation}
p\!\left(e_{j_a}^{(a)} \middle| \lambda_{\alpha_a}^{(a)}\right)
=
\frac{
\Tr\!\left[\Pi_{j_a}^{(a)}P_{\alpha_a}^{(a)}\rho_a P_{\alpha_a}^{(a)}\right]
}{
\Tr\!\left[P_{\alpha_a}^{(a)}\rho_a\right]
}.
\label{eq:conditional_energy_given_state}
\end{equation}
If $p(\alpha_1,\ldots,\alpha_5)$ denotes the joint probability distribution associated with the state-projector outcomes, then the inferred energy-path distribution is
\begin{equation}
p_{\rm DBN}(j_1,\ldots,j_5)
=
\sum_{\alpha_1,\ldots,\alpha_5}
p(\alpha_1,\ldots,\alpha_5)
\prod_{a=1}^{5}
p\!\left(e_{j_a}^{(a)} \middle| \lambda_{\alpha_a}^{(a)}\right).
\label{eq:pdbn_path}
\end{equation}
This is the central probabilistic object in our fluctuation analysis. Each inferred energy history
\begin{equation}
\mathbf{j}=(j_1,j_2,j_3,j_4,j_5)
\end{equation}
is assigned stochastic work and heat contributions through the corner-energy differences,
\begin{align}
W_{c\to h}(\mathbf{j}) &= e_{j_2}^{(2)}-e_{j_1}^{(1)},
\\
Q_h(\mathbf{j}) &= e_{j_3}^{(3)}-e_{j_2}^{(2)},
\\
W_{h\to c}(\mathbf{j}) &= e_{j_4}^{(4)}-e_{j_3}^{(3)},
\\
Q_c(\mathbf{j}) &= e_{j_5}^{(5)}-e_{j_4}^{(4)},
\end{align}
so that the total stochastic work is
\begin{equation}
W(\mathbf{j})=W_{c\to h}(\mathbf{j})+W_{h\to c}(\mathbf{j}).
\label{eq:w_path}
\end{equation}
For the inferred DBN energy histories, branchwise energetic additivity is exact by construction. However, the pathwise cyclic identity takes the form
\begin{equation}
W(\mathbf j)+Q_h(\mathbf j)+Q_c(\mathbf j)=\Delta E_{\rm cyc}(\mathbf j)
=e_{j_5}^{(5)}-e_{j_1}^{(1)},
\end{equation}
which does not, in general, vanish trajectory by trajectory. Instead, strict cycle closure is recovered only at the ensemble level, yielding the standard first law in the stationary regime,
\begin{equation}
\langle W\rangle+\langle Q_h\rangle+\langle Q_c\rangle
=
\Tr[H_c(\rho_5-\rho_1)]
\simeq 0,
\end{equation}
up to the numerical tolerance used to determine the limit cycle. This distinction is consistent with the DBN literature, where the minimally invasive character of the protocol refers to averaged measured states and averaged energetics, rather than to an exact pathwise cyclic constraint. More generally, coherence-preserving inference schemes do not, in general, admit the same strictly classical trajectory interpretation as fully projective constructions, since the preservation of coherence fundamentally allows for interfering quantum paths and histories \cite{Hofer2016,Solinas2016,Miller2017}. In this sense, the absence of exact pathwise cycle closure should not be viewed as an inconsistency of the present DBN construction, but as part of the fundamental trade-off involved in retaining coherence-sensitive fluctuation statistics while preserving the unperturbed average energetics \cite{Allahverdyan2014,Perarnau2017,Lostaglio2023,Hovhannisyan2024}.

\subsection{Fluctuation observables}
\label{sec:fluct_obs}

Once the DBN path distribution $p_{\rm DBN}(j_1,\dots,j_5)$ is constructed, any cycle observable $X(\mathbf j)$ can be averaged as
\begin{equation}
\langle X\rangle = \sum_{j_1,\ldots,j_5} X(\mathbf j)\,p_{\rm DBN}(j_1,\ldots,j_5),
\label{eq:X_average_dbn}
\end{equation}
with variance
\begin{equation}
\text{Var}(X)=\langle X^2\rangle-\langle X\rangle^2.
\label{eq:X_variance_dbn}
\end{equation}
In particular, we evaluate the fluctuations of work and heat through $\text{Var}(W)$ and $\text{Var}(Q_{i=c,h})$ and the corresponding standard deviations $\sigma(W)=\sqrt{\text{Var}(W)}$ and $\sigma(Q_{i=c,h})=\sqrt{\text{Var}(Q_{i=c,h})}$.

To quantify the strength of fluctuations relative to the mean signal, we introduce the relative fluctuation of a generic observable $O$ as \cite{Xiao2023}
\begin{equation}
\phi_O\equiv \frac{\sigma(O)}{|\langle O\rangle|},
\label{eq:phiO_def}
\end{equation}
whenever $\langle O\rangle\neq 0$. The corresponding reliability \cite{Jaseem2023} is the inverse quantity ,
\begin{equation}
R_O\equiv \frac{|\langle O\rangle|}{\sigma(O)}=\phi_O^{-1}.
\label{eq:RO_def}
\end{equation}

In the present work, the heatmaps display the relative fluctuations $\phi_O$, while the reliabilities $R_O$ are used only when we wish to discuss consistency in the strict sense of mean output divided by its standard deviation. The literal stochastic efficiency
\begin{equation}
\eta_{\rm st}(\mathbf j)=-\frac{W(\mathbf j)}{Q_h(\mathbf j)}
\label{eq:eta_literal}
\end{equation}
is not used as a primary map variable because it becomes singular whenever $Q_h(\mathbf j)\to 0$ and develops the well-known broad tails of stochastic efficiency statistics \cite{Denzler2020,Fei2022}. These singular events are physical features of the ratio distribution, not numerical errors. Since the present work focuses on robust parameter-space diagnostics rather than on deriving an efficiency fluctuation theorem, we instead use the regularized normalized-work proxy
\begin{equation}
\eta_{\rm reg}(\mathbf j)\equiv -\frac{W(\mathbf j)}{\langle Q_h\rangle}.
\label{eq:eta_reg_def}
\end{equation}
Its average and variance are
\begin{equation}
\langle \eta_{\rm reg}\rangle = -\frac{\langle W\rangle}{\langle Q_h\rangle}\equiv \eta,
\qquad
\Var(\eta_{\rm reg})=\frac{\Var(W)}{\langle Q_h\rangle^2},
\label{eq:eta_reg_moments}
\end{equation}
so that
\begin{equation}
\sigma(\eta_{\rm reg})=\frac{\sigma(W)}{\langle Q_h\rangle}.
\label{eq:eta_reg_sigma}
\end{equation}
Thus $\eta_{\rm reg}(\mathbf j)$ is not a genuine trajectory efficiency in the same sense as Eq.~\eqref{eq:eta_literal}; it is a finite normalized-work diagnostic that retains the correct thermodynamic mean efficiency $\eta$ while replacing the fluctuating heat denominator by the mean hot heat. It therefore cannot be used to establish standard efficiency fluctuation relations, and it does not probe the singular tails or sign-changing heat events of the full ratio distribution. A full stochastic-efficiency analysis would require the joint DBN distribution of $(W,Q_h)$ and a separate treatment of the singular sectors. Accordingly, all statements involving $\eta_{\rm reg}$ should be read with this limitation.

The corresponding relative fluctuation of the regularized proxy is
\begin{equation}
\phi_{\eta_{\rm reg}}
\equiv
\frac{\sigma(\eta_{\rm reg})}{\eta}
=
\frac{\sigma(W)}{|\langle W\rangle|}
=
\phi_P,
\label{eq:phi_eta_def}
\end{equation}
where the last equality follows because the cycle time is deterministic. Hence $\phi_{\eta_{\rm reg}}$ is not an independent efficiency-noise observable; it is identical to the relative work or relative power fluctuation. For this reason, the main stability discussion below is phrased primarily in terms of $\sigma(W)$ and $\phi_P$, while $\sigma(\eta_{\rm reg})$ is retained only as a normalized-work fluctuation scale.

Since the cycle time $\tau_{\rm cyc}$ is deterministic, the power fluctuations follow directly from the work fluctuations. Defining the stochastic power along a path as
\begin{equation}
P(\mathbf j)=-\frac{W(\mathbf j)}{\tau_{\rm cyc}},
\label{eq:P_path}
\end{equation}
we obtain
\begin{equation}
\text{Var}(P)=\frac{\text{Var}(W)}{\tau_{\rm cyc}^2}, \qquad \sigma(P)=\frac{\sigma(W)}{\tau_{\rm cyc}}.
\label{eq:P_var_sigma}
\end{equation}
Hence the relative power fluctuation is
\begin{equation}
\phi_P \equiv \frac{\sigma(P)}{|\langle P\rangle|} = \frac{\sigma(W)}{|\langle W\rangle|},
\label{eq:power_rel}
\end{equation}
while the corresponding power reliability is
\begin{equation}
R_P=\phi_P^{-1}=\frac{|\langle P\rangle|}{\sigma(P)}.
\label{eq:power_reliability}
\end{equation}
In the analysis below, the main fluctuation diagnostics are therefore $\sigma(W)$, $\sigma(\eta_{\rm reg})$, and $\phi_P$; the reliability $R_P$ is introduced only when reliability is discussed explicitly.

\section{Full-thermalization benchmark with a population-inverted hot channel}
\label{sec:idealnegative}

\begin{figure*}[t!]
	\centering
	\includegraphics[width=\textwidth]{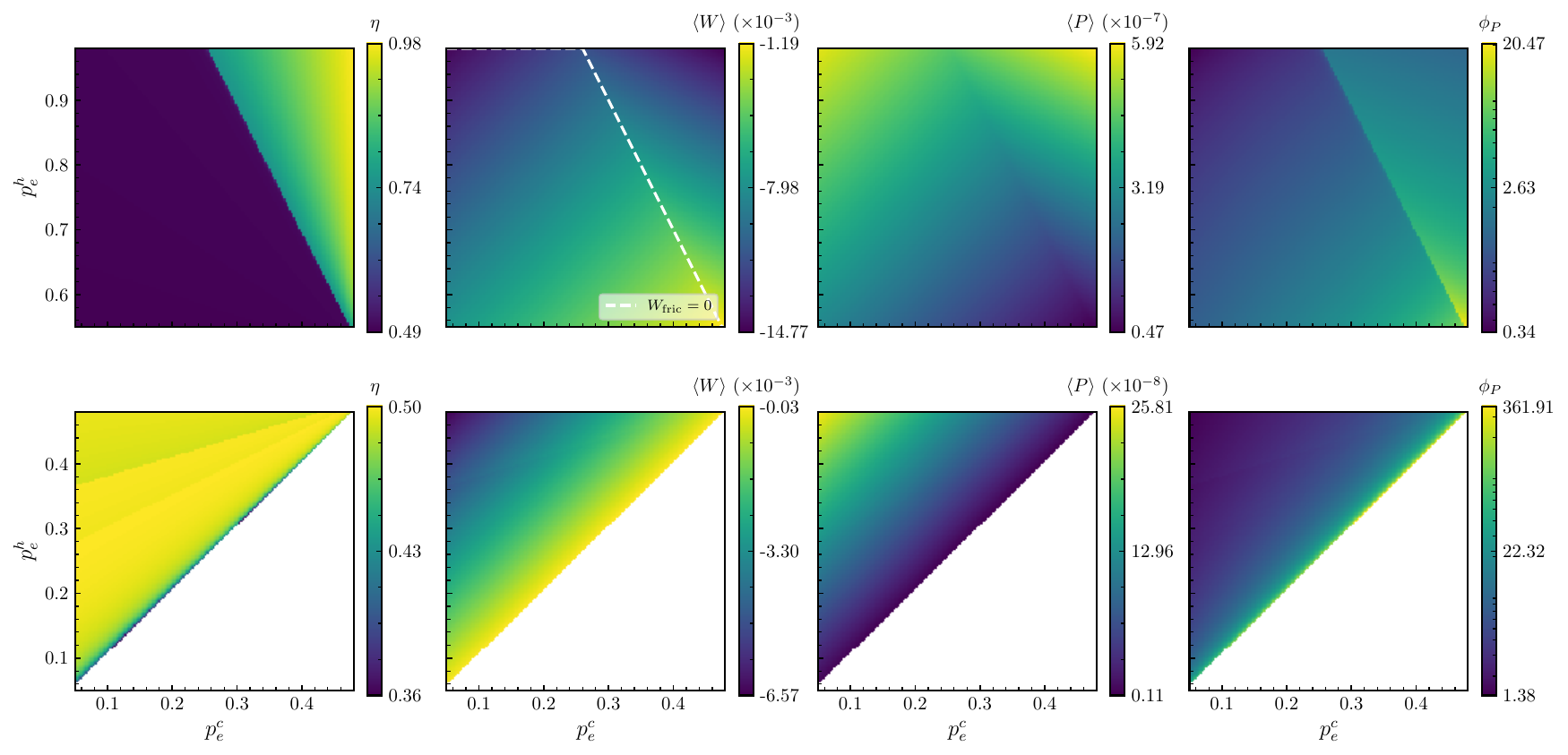}
	\caption{
		Comparative fluctuation landscape in the ideal-reset regime. The top row corresponds to the population-inverted hot channel, equivalently the effective negative-temperature regime ($T_h<0$, $p_e^h>1/2$), while the bottom row corresponds to the positive-temperature reference case ($T_h>0$, $p_e^h<1/2$). The columns show the efficiency $\eta_{\rm reg}$, mean work $\langle W\rangle$, output power $P$, and relative power fluctuation $\phi_P$ in the population plane $(p_e^c,p_e^h)$, all evaluated at the driving time $\tau_{\rm drive}^{\star}$ that maximizes the output power for each population pair. The dashed curve in the inverted-regime work panel marks $W_{\rm fric}=0$. White regions indicate points outside the engine regime, namely parameter sectors in which the cycle does not satisfy the engine conditions $\langle W\rangle<0$ and $\langle Q_h\rangle>0$. Within the reduced inverted-channel model, the engine develops a broader gross-performance window, with extracted work and power enhanced by roughly a factor of two relative to the best positive-temperature region, while also supporting a lower-relative-noise stability sector.
	}
	\label{fig:combined_heatmaps}
\end{figure*}

We first analyze the full-thermalization limit, which provides a controlled benchmark for separating the effect of population inversion from incomplete bath relaxation. In this limit, the hot and cold isochores are long enough to reset the working medium close to the corresponding stationary states after each dissipative stroke. The cycle therefore has no inter-cycle memory associated with incomplete thermalization, while still retaining the coherence generated during the finite-time unitary branches connecting the noncommuting Hamiltonians $H_c$ and $H_h$.

This benchmark is useful for two reasons. First, it connects the present DBN analysis to previous studies of qubit Otto engines with effective negative-temperature reservoirs, where population inversion was shown to enhance mean work extraction and finite-time performance \cite{Xi2017,deAssis2019,Nettersheim2022,desousa2024}. Second, it establishes a clean reference against which the genuinely finite-time regime can be interpreted. Any stability or fluctuation structure found here arises from the combination of coherent driving and the inverted stationary populations, not from incomplete thermalization or memory across cycles.

In the numerical implementation, the full-thermalization condition is enforced by choosing the hot- and cold-bath durations long enough that representative initial states relax to the corresponding stationary states within the imposed tolerances. This convergence is monitored through both the trace distance to the stationary density matrix and the residual coherence in the instantaneous energy basis. The resulting limit should be read as an ideal reset benchmark of the reduced engine model. Since the inverted hot channel is actively maintained, the performance reported in this section is a gross working-medium performance and does not include the external resource cost required to prepare or stabilize the inversion.

To construct the benchmark maps, we scan the population plane $(p_e^c,p_e^h)$ and, for each point, determine the driving time $\tau_{\rm drive}^{\star}$ that maximizes the output power. The remaining observables are then evaluated at that same power-optimal driving time. Figure~\ref{fig:combined_heatmaps} compares the resulting ideal-reset landscapes for positive-temperature and population-inverted hot channels. The efficiency shown in the first column is therefore the efficiency at $\tau_{\rm drive}^{\star}$, not the maximum efficiency over all driving times.

\subsection{Performance and stability trade-off}

Figure~\ref{fig:combined_heatmaps} shows that the population-inverted reduced model opens a broader operating window than the positive-temperature reference. In the positive-temperature case, the power-optimized landscape remains close to the standard Otto expectation, with efficiencies bounded near $1-\omega_c/\omega_h\simeq 0.5$ and comparatively modest work extraction and power. In the inverted case, the high-performance region expands substantially: the extracted work and output power increase, and the efficiency approaches unity over a broad sector of the accessible population plane. The white regions indicate points outside the engine regime, where the cycle does not simultaneously satisfy $\langle W\rangle<0$ and $\langle Q_h\rangle>0$.

Representative operating points are summarized in Table~\ref{tab:ideal_reset_comparison}. At the best positive-temperature operating point, the engine reaches $|\langle W\rangle|=6.57\times10^{-3}$, $P=2.58\times10^{-7}$, and $\eta_{\rm reg}=0.493$. At the corresponding inverted-channel performance optimum, these values become $|\langle W\rangle|=14.8\times10^{-3}$, $P=5.92\times10^{-7}$, and $\eta_{\rm reg}=0.979$. Thus, within the reduced model, population inversion enhances both extracted work and output power by roughly a factor of $2.3$. This should be understood as a gross working-medium enhancement, not as a net device-level gain after reservoir-preparation costs.

The fluctuation structure is equally important. The inverted landscape contains a stability sector in which the engine retains nearly the same useful output as at the performance optimum, but with a much smaller relative power fluctuation. At the stability point listed in Table~\ref{tab:ideal_reset_comparison}, the output power remains comparable to that of the inverted performance optimum, while the relative power fluctuation drops to $\phi_P=0.338$, compared with $\phi_P=1.38$ at the best positive-temperature point. The regularized efficiency-proxy fluctuation follows the same trend, but, as discussed above, its relative fluctuation is algebraically identical to $\phi_P$ and should not be interpreted as an independent stochastic-efficiency measure.

This comparison shows that the ideal-reset inverted channel improves not only mean output but also the accessible performance--stability trade-off within the reduced working-medium dynamics. Absolute fluctuation minimization alone would be misleading, since low-output regions can always appear stable. The relevant diagnostic is the joint behavior of extracted work, power, efficiency, and relative power noise.

\begin{table*}[t]
	\caption{
		Comparison of representative operating points in the ideal-reset regime for positive-temperature ($T_h>0$) and effective negative-temperature ($T_h<0$) hot channels. All values are obtained from the DBN analysis. Acronyms: MP = maximum power, ME = maximum efficiency, and MS = minimum relative power fluctuation. The quoted efficiency-fluctuation scale is the regularized normalized-work quantity $\sigma(\eta_{\rm reg})$, not the variance of the literal stochastic ratio $-W/Q_h$.
	}
	\label{tab:ideal_reset_comparison}
	\centering
	\begin{ruledtabular}
		\begin{tabular}{lccccccccc}
			Regime & Point & $p_e^c$ & $p_e^h$
			& $|\langle W\rangle|$ $(10^{-3})$
			& $\sigma(W)$ $(10^{-3})$
			& $\eta_{\rm reg}$
			& $\sigma(\eta_{\rm reg})$
			& $P$ $(10^{-7})$
			& $\phi_P$ \\
			\hline
			$T_h>0$ & MP = MS
			& 0.050 & 0.480
			& 6.57 & 9.05 & 0.493 & 0.679 & 2.58 & 1.38 \\
			$T_h>0$ & ME
			& 0.163 & 0.333
			& 2.65 & 9.46 & 0.497 & 1.77 & 1.00 & 3.57 \\
			\hline
			$T_h<0$ & MP = ME
			& 0.480 & 0.980
			& 14.8 & 19.7 & 0.979 & 1.31 & 5.92 & 1.34 \\
			$T_h<0$ & MS
			& 0.050 & 0.980
			& 14.5 & 4.91 & 0.500 & 0.169 & 5.71 & 0.338 \\
		\end{tabular}
	\end{ruledtabular}
\end{table*}

\subsection{Friction structure and physical interpretation}

The dashed curve in the upper work panel of Fig.~\ref{fig:combined_heatmaps} marks the condition $W_{\rm fric}=0$, namely the boundary where the friction contribution changes sign. We use this curve as an organizing diagnostic rather than as an optimization criterion. It connects the present fluctuation analysis with the thermodynamic interpretation of Ref.~\cite{desousa2024}, where the inverted Otto engine was analyzed in terms of entropy production and friction work.

For the present qubit protocol, the friction contribution becomes negative when
\begin{equation}
	\omega_h(1-2p_e^c)+\omega_c(1-2p_e^h)<0.
	\label{eq:negative_friction_condition}
\end{equation}
Equivalently, the friction-zero boundary is
\begin{equation}
	p_e^h=
	\frac{1}{2}
	\left[
	1+\frac{\omega_h}{\omega_c}(1-2p_e^c)
	\right].
	\label{eq:friction_zero_line}
\end{equation}
Thus the onset of negative friction is controlled jointly by the population inversion and the spectral asymmetry of the Otto cycle.

Physically, this sign change reflects the fact that nonadiabatic transitions do not have the same thermodynamic role in an inverted population distribution as they do in a passive positive-temperature distribution. In a passive engine, nonadiabatic excitations generally increase the energetic cost of the unitary strokes and reduce useful output. In the inverted case, the population ordering is reversed, so the same transition structure can contribute constructively to the work balance. The term \emph{negative friction} should therefore be understood as a sign reversal of the nonadiabatic work correction relative to the adiabatic reference, not as an absence of irreversibility or as a violation of the second law.

In the positive-temperature regime, nonadiabatic friction is purely detrimental: it increases the energetic cost of the cycle and reduces useful work extraction. In the inverted regime, by contrast, there exists a sector where the friction contribution becomes negative, so nonadiabaticity can assist gross work extraction within the reduced model. The high-performance region in Fig.~\ref{fig:combined_heatmaps} lies on the side of the population plane where this mechanism is active.

The friction-zero line should not be overinterpreted. It does not determine the optimal operating point and does not replace the fluctuation analysis. Its role is to explain why the inverted-channel landscape differs qualitatively from the positive-temperature reference. The main conclusion of the ideal-reset benchmark is that population inversion can generate a broader gross-performance window and a distinct low-relative-noise sector before incomplete-thermalization effects are introduced.

The ideal-reset benchmark therefore serves as the reference point for the finite-time analysis. It establishes what is already possible in the reduced model when the isochores act as effective reset strokes, and it separates this behavior from the additional memory and coherence-survival effects that appear when the bath-contact times are finite. We now turn to that regime and ask how the inverted-channel operating landscape is reorganized once the cycle is no longer reset exactly to the reservoir stationary states after each round.

\section{Finite-time operating landscapes}
\label{sec:finite_time_landscapes}

We now move beyond the ideal-reset benchmark and consider finite-duration isochores. In this regime, the hot and cold strokes no longer reset the working medium exactly to their stationary states, so incomplete thermalization and inter-stroke memory can reshape both the mean output and the fluctuation structure. This is the regime most relevant for finite-time operation in engineered quantum devices, because increasing the cycle rate generally competes with thermalization quality and output stability.

To keep the parameter space transparent, we impose symmetric bath-contact times,
\begin{equation}
	\tau_h=\tau_c\equiv\tau_{\rm iso},
\end{equation}
and scan the plane $(\tau_{\rm drive},\tau_{\rm iso})$. The resulting maps should be read as reduced-model operating landscapes. They do not include the external cost of maintaining the inverted channel, but they identify where active inversion, finite-time driving, relative power noise, and coherence preservation become simultaneously relevant.

\subsection{Finite-time population-inverted engine}
\label{sec:finite_negative}

We first analyze the finite-time landscape of the population-inverted reduced model. Throughout this subsection, the bath populations are fixed at
\begin{equation}
	p_e^c=0.40,
	\qquad
	p_e^h=0.80,
	\label{eq:finite_time_populations}
\end{equation}
so that the hot channel is inverted while the cold channel remains non-inverted. For the frequency ratio used in this work, this point lies in the sector where the friction contribution can become negative, meaning that coherent nonadiabatic driving can assist gross work extraction rather than acting only as a loss mechanism; see Sec.~\ref{sec:idealnegative} and Ref.~\cite{desousa2024}.

\begin{figure*}[t!]
	\centering
	\includegraphics[width=1.05\textwidth]{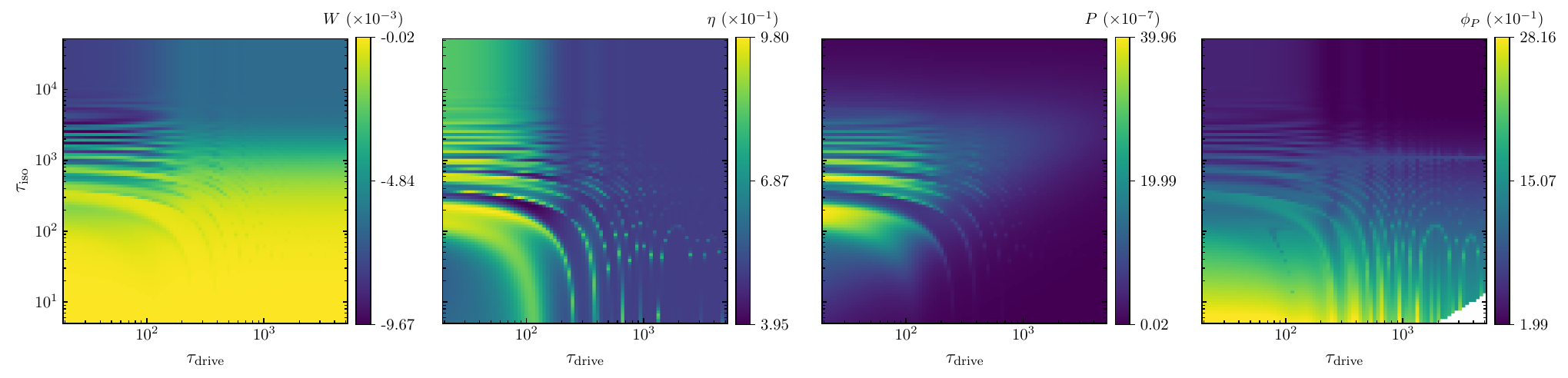}
	\caption{
		Finite-time reduced population-inverted engine in the $(\tau_{\rm drive},\tau_{\rm iso})$ plane. Heatmaps show the mean work $\langle W\rangle$, efficiency $\eta_{\rm reg}$, output power $P$, and relative power fluctuation $\phi_P=\sigma(P)/|\langle P\rangle|$. White regions correspond to points outside the engine regime. The finite-time landscape separates into distinct operational sectors: a short-time high-power sector, a narrow high-efficiency but strongly fluctuating sector, and a long-time low-relative-noise sector approaching the reset-like regime. The maps therefore provide design diagnostics rather than a single universal optimum.
	}
	\label{fig:finite_negative_landscape}
\end{figure*}

Figure~\ref{fig:finite_negative_landscape} and Table~\ref{tab:finite_negative_points} show that finite bath-contact times reorganize, rather than eliminate, the inverted-channel enhancement found in the reset benchmark. The maximum-power point reaches $P=40.0\times10^{-7}$, nearly seven times larger than the reset-limit value $5.92\times10^{-7}$ in Table~\ref{tab:ideal_reset_comparison}. This optimum occurs at short driving time, $\tau_{\rm drive}=20$, and intermediate isochore duration, $\tau_{\rm iso}\simeq575$, forming a pronounced high-power lobe. Within the reduced working-medium dynamics, finite-time operation therefore amplifies the gross output relative to the ideal-reset benchmark.

The maximum-efficiency sector is qualitatively different. Near-unit efficiency survives, with $\eta_{\rm reg}=0.980$ at $(\tau_{\rm drive},\tau_{\rm iso})=(35.0,206.6)$, but this point is accompanied by very large relative power fluctuations, $\phi_P=14.3$. Thus high efficiency is not by itself a useful design criterion in the finite-time regime. Incomplete thermalization breaks the near coincidence between high efficiency and good stability observed in the reset benchmark, producing instead a narrow and strongly fluctuating high-efficiency branch.

The opposite limit reveals a long-time stability sector. The minimum relative power fluctuation, $\phi_P=1.58$, occurs at $(\tau_{\rm drive},\tau_{\rm iso})=(2666,16373)$. This point has much smaller power, $P=1.65\times10^{-7}$, but retains a sizable extracted work, $|\langle W\rangle|=6.28\times10^{-3}$, and relaxes toward $\eta\simeq0.50$. It therefore behaves as a reset-like reliability plateau: less powerful, but substantially less noisy. Since the heatmaps display $\phi_P$ rather than $R_P=\phi_P^{-1}$, low values of $\phi_P$ correspond to high relative reliability.

For a practical reduced-model ranking, we introduce the empirical power--stability score
\begin{equation}
	\mathcal{F}\equiv \frac{P}{\phi_P}
	=P R_P
	=
	\frac{P^2}{\sigma(P)}.
	\label{eq:power_stability_score}
\end{equation}
This quantity is not a universal thermodynamic bound; it is a pragmatic diagnostic that rewards large output while penalizing large relative power noise. Maximizing $\mathcal{F}$ selects the trade-off point in Table~\ref{tab:finite_negative_points}, located at $(\tau_{\rm drive},\tau_{\rm iso})=(46.3,1328)$. This point retains a large fraction of the finite-time gross power enhancement, $P=30.2\times10^{-7}$, while reducing the relative power fluctuation to $\phi_P=2.43$, compared with $\phi_P=5.00$ at the maximum-power point.

From a design perspective, the important message is that the finite-time landscape does not possess a single universal optimum. Instead, it separates into operational sectors. The short-time sector is attractive for large gross power but is fluctuation limited. The high-efficiency sector is narrow and strongly noisy. The long-time sector sacrifices power but approaches a more reliable reset-like regime. The trade-off point selected by $\mathcal{F}$ is therefore not a fundamental optimum, but a useful diagnostic for identifying operating regions that retain significant output while suppressing relative power noise.

\begin{table*}[t]
	\caption{
		Representative operating points of the finite-time population-inverted DBN engine in the $(\tau_{\rm drive},\tau_{\rm iso})$ plane.
		Acronyms: MW = maximum extracted work, MP = maximum power, ME = maximum efficiency, MS = minimum relative power fluctuation, and TO = trade-off operating point selected by the empirical power--stability score $\mathcal{F}=P/\phi_P=P R_P$, where $\phi_P=\sigma(P)/|\langle P\rangle|$ and $R_P=|\langle P\rangle|/\sigma(P)$.
	}
	\label{tab:finite_negative_points}
	\centering
	\begin{ruledtabular}
		\begin{tabular}{lccccccc}
			Point & $\tau_{\rm drive}$ & $\tau_{\rm iso}$ & $|\langle W\rangle|$ $(10^{-3})$ & $\sigma(W)$ $(10^{-3})$ & $\eta_{\rm reg}$ & $P$ $(10^{-7})$ & $\phi_P$ \\
			\hline
			MW   & 30.4   & 2547   & 9.67 & 20.4 & 0.879 & 18.8 & 2.11 \\
			MP   & 20.0   & 574.9  & 4.75 & 23.8 & 0.933 & 40.0 & 5.00 \\
			ME   & 35.0   & 206.6  & 1.66 & 23.6 & 0.980 & 34.3 & 14.3 \\
			MS   & 2666   & 16373  & 6.28 & 9.94 & 0.500 & 1.65 & 1.58 \\
			TO   & 46.3   & 1328   & 8.31 & 20.2 & 0.883 & 30.2 & 2.43 \\
		\end{tabular}
	\end{ruledtabular}
\end{table*}

\subsection{Coherence-sensitive structure of the finite-time regime}
\label{sec:coherence_finite}

We next ask whether the operational sectors identified above are correlated with residual coherence in the working medium. This question is particularly relevant for the DBN analysis, because DBN is intended to reconstruct the fluctuation statistics of the coherent, unmeasured cycle, whereas TPM would dephase the system in the energy basis. The coherence map therefore plays a diagnostic role: it indicates where a coherence-preserving reconstruction is expected to matter most.

We quantify the residual coherence after the hot isochore using the relative-entropy coherence in the energy basis of $H_h$ \cite{Streltsov2017,Baumgratz2014},
\begin{equation}
	C_{\mathrm{rel}}(\rho_3;H_h)
	=
	S\!\left(\Delta_{H_h}[\rho_3]\right)-S(\rho_3),
	\label{eq:Crel_hot}
\end{equation}
where $\rho_3$ is the state at the end of the hot stroke, $S(\rho)=-\Tr(\rho\ln\rho)$, and $\Delta_{H_h}$ denotes complete dephasing in the eigenbasis of $H_h$. We use $C_{\mathrm{rel}}$ because it has a direct information-theoretic interpretation as the entropy increase under energy-basis dephasing. Other coherence measures, such as the $l_1$ norm, could be used as complementary diagnostics; the present conclusions should therefore be read as statements about the structure of this energy-basis coherence landscape, not as a claim of measure-independent coherence quantification.

The full coherence landscapes are shown in Appendix~\ref{app:coherence}, Fig.~\ref{fig:coherence_landscape_comparison}, and representative values are listed in Table~\ref{tab:coherence_summary}. The absolute coherence scale remains modest compared with the qubit upper bound $\ln 2$. The relevant observation is therefore not that the engine operates with a large coherence resource, but that the alignment between coherence and useful operating sectors is strongly regime dependent.

In the positive-temperature reference, the maximum post-hot-bath coherence is clearly separated from the best thermodynamic sectors. Although the global coherence maximum reaches $1.55\times10^{-2}$, the coherence at the maximum-power and maximum-efficiency points is only $1.29\times10^{-4}$ and $1.71\times10^{-5}$, respectively. Thus the best-performing positive-temperature operating points lie in an almost fully decohered region of the finite-time landscape.

The population-inverted case shows a different structure. Its global coherence maximum is larger, $C_{\mathrm{rel}}^{\max}\simeq3.56\times10^{-2}$, and the maximum-efficiency point occurs at $C_{\mathrm{rel}}\simeq2.53\times10^{-2}$, close to the dominant coherence ridge. The maximum-power point has a smaller but still non-negligible value, $C_{\mathrm{rel}}\simeq5.73\times10^{-3}$. Hence the inverted high-efficiency sector remains structurally aligned with residual post-hot-bath coherence, even though the absolute coherence remains modest on the qubit scale---roughly two orders of magnitude below the theoretical maximum of $\ln 2 \approx 0.693$.

This comparison gives the coherence map its operational meaning.
We stress that the observed alignment reflects a structural correlation, not a direct causal enhancement;
the underlying physical mechanism is the nonadiabatic inner friction that assists work extraction in the inverted regime
through Eq.~\eqref{eq:negative_friction_condition}.
The relevance of the map lies in identifying where a coherence‑preserving reconstruction becomes operationally significant.
In the positive‑temperature reference, the optimal sectors are nearly diagonal in the hot energy basis,
so a projective TPM description is less disruptive.
In the inverted case, the high‑efficiency branch remains close to the coherence ridge,
and as demonstrated in Appendix~\ref{app:dbn_tpm_benchmark}, DBN and TPM predictions diverge quantitatively
in precisely these short‑time, coherence‑rich sectors, shifting the apparent optimal operating boundaries.
Therefore, the coherence‑sensitive result is structural rather than resource‑magnitude based,
and it does not imply that larger coherence universally improves performance.
Testing the same alignment with $l_1$ coherence or at other cycle corners would provide a useful robustness check,
left for future work.

\begin{table}[t]
	\caption{
		Relative-entropy coherence after the hot bath, $C_{\mathrm{rel}}(\rho_3;H_h)$, evaluated at representative operating points of the finite-time DBN engine. The positive-temperature row corresponds to the reference finite-time scan used as the baseline for comparison. The table highlights the alignment, or lack of alignment, between post-hot-bath coherence and the maximum-power and maximum-efficiency sectors.
	}
	\label{tab:coherence_summary}
	\centering
	\begin{ruledtabular}
		\begin{tabular}{lccc}
			Case & $C_{\mathrm{rel}}^{\max}$ & at MP & at ME \\
			\hline
			Positive, $T_h>0$ & \(1.55\times10^{-2}\) & $1.29\times10^{-4}$ & $1.71\times10^{-5}$ \\
			Negative, $T_h<0$ & $3.56\times10^{-2}$ & $5.73\times10^{-3}$ & $2.53\times10^{-2}$ \\
		\end{tabular}
	\end{ruledtabular}
\end{table}

\section{Discussion}
\label{sec:discussion}

The results above provide a reduced-model operating map for a finite-time qubit Otto engine coupled to an actively maintained population-inverted hot channel. The main physical message is that population inversion, finite-time driving, fluctuation stability, and coherence preservation cannot be assessed independently. In the ideal-reset benchmark, the inverted channel broadens the gross-performance window and opens a low-relative-noise stability sector. Once the isochores are made finite, this behavior is reorganized into distinct operating sectors: a short-time high-power branch, a narrow high-efficiency but strongly fluctuating branch, and a long-time low-relative-noise branch approaching the reset-like regime.

This separation of sectors is the most useful design-level outcome of the analysis. It shows that optimizing an engineered qubit engine by a single scalar quantity can be misleading. Maximum efficiency does not coincide with maximum stability in the finite-time regime, and maximum power comes with a non-negligible relative-noise cost. The empirical score $\mathcal{F}=P/\phi_P$ is therefore not introduced as a universal thermodynamic optimum, but as a practical diagnostic for identifying operating regions that retain substantial gross output while suppressing relative power fluctuations. In this sense, the finite-time maps should be read as screening tools for future platform-specific implementations.

The interpretation of the inverted channel must remain resource-qualified. The hot stroke is described by a Markovian channel with an inverted stationary population satisfying the rate ratio in Eq.~\eqref{eq:rate_ratio}. The finite-time working point $p_e^h=0.80$ requires $\Gamma_\uparrow/\Gamma_\downarrow=4$, while the strongly inverted ideal-reset point $p_e^h=0.98$ requires $\Gamma_\uparrow/\Gamma_\downarrow=49$. These values make clear that the hot channel is an active reservoir-engineering primitive rather than an ordinary passive heat bath. A complete device-level analysis would need to include the work, entropy production, leakage, heating, and possible finite-resource effects associated with preparing or continuously maintaining this inversion.

If the external pump or preparation work per cycle is denoted by $W_{\rm pump}\geq 0$, a simple resource-accounted efficiency would have the schematic form
\begin{equation}
    \eta_{\rm tot}
    =
    \frac{-\langle W\rangle}
    {\langle Q_h\rangle+W_{\rm pump}}.
    \label{eq:eta_total_resource}
\end{equation}
More generally, the appropriate device-level figure of merit should be formulated in terms of the exergy supplied by the active reservoir. Since the present work does not model the pump or synthetic-reservoir mechanism explicitly, Eq.~\eqref{eq:eta_total_resource} is not evaluated numerically. 
This expression also provides a simple threshold interpretation. If
$\eta_{\rm gross}=-\langle W\rangle/\langle Q_h\rangle$
denotes the efficiency of the reduced working-medium model, then maintaining a target net efficiency $\eta_{\rm target}<\eta_{\rm gross}$ would require
\begin{equation}
\frac{W_{\rm pump}}{\langle Q_h\rangle}
<
\frac{\eta_{\rm gross}}{\eta_{\rm target}}-1 .
\label{eq:pump_threshold}
\end{equation}
For example, taking $\eta_{\rm target}$ to be the positive-temperature Otto efficiency gives a direct dimensionless bound on the admissible pump overhead. This bound is not a microscopic resource accounting; it is only a screening criterion indicating how much external overhead the gross inverted-channel enhancement could tolerate.

The reported efficiencies and powers are therefore gross working-medium quantities. They identify where the reduced engine dynamics is favorable, but they do not by themselves establish net device-level superiority over a fully resource-accounted positive-temperature engine.

The same caution applies to the efficiency-fluctuation language. The regularized variable $\eta_{\rm reg}(j)=-W(j)/\langle Q_h\rangle$ is a normalized stochastic work output, not the literal trajectory efficiency $-W(j)/Q_h(j)$. It removes the singular denominator responsible for the broad tails of genuine stochastic-efficiency statistics and therefore cannot be used to establish standard efficiency fluctuation relations. Its role in this work is diagnostic: it provides a stable normalized scale for comparing large parameter scans. The independent stability information is primarily contained in $\sigma(W)$ and $\phi_P$, while the relative fluctuation of $\eta_{\rm reg}$ is algebraically identical to $\phi_P$.

The coherence analysis also supports a deliberately structural interpretation. The absolute values of the relative-entropy coherence remain modest compared with the qubit upper bound $\ln 2$, so the result should not be read as evidence for a large coherence resource. Rather, the important observation is that the alignment between coherence and useful operating sectors is regime dependent. In the positive-temperature reference, the maximum-power and maximum-efficiency sectors lie in an almost decohered region. In the population-inverted case, the high-efficiency branch remains close to the dominant post-hot-bath coherence ridge. Thus coherence is not universally beneficial, but it identifies where a coherence-preserving reconstruction becomes operationally relevant.

This point also clarifies the role of the DBN framework. DBN is not a universal replacement for TPM and is not a single-shot backaction-free measurement protocol. It is an ensemble-level reconstruction of the coherent, unmeasured cycle. An experimental implementation would require repeated preparation of the stationary cycle, tomography of the five corner states, and estimation of the conditional probabilities entering the Bayesian network. This overhead is greater than in a direct TPM experiment, and the present work does not include sampling-error or tomography-noise propagation. The advantage of DBN in the present context is that it targets the fluctuation statistics of the untouched coherent dynamics, whereas TPM characterizes the explicitly measured and dephased engine.

Taken together, these qualifications define the technological scope of the present study. The manuscript does not propose a complete powered heat-engine device. Instead, it provides a reduced-model diagnostic framework for engineered qubit thermal machines with active reservoirs. The finite-time maps identify which operating sectors are promising from the standpoint of gross output, relative-noise suppression, and coherence-sensitive reconstruction. Natural next steps are therefore platform-specific: implementing an explicit pump or synthetic-reservoir model, propagating tomography errors through the DBN reconstruction, and testing whether the gross operating sectors identified here survive after full resource accounting.

\section{Conclusion}
\label{sec:conclusion}

We have developed a reduced-model fluctuation diagnostic for a finite-time qubit Otto engine coupled to an actively maintained population-inverted hot channel. The analysis combines finite-time driving, incomplete thermalization, coherence-sensitive reconstruction, and fluctuation stability within a single operating-map framework. The dynamic Bayesian network construction was used as an ensemble-level reconstruction of the unmeasured coherent cycle, while TPM was interpreted as the corresponding protocol for an explicitly measured and dephased engine.

Within this reduced working-medium model, population inversion substantially modifies the operating landscape. In the full-thermalization benchmark, the inverted channel enhances gross extracted work and output power while opening a low-relative-noise stability sector. When the hot and cold isochores are made finite, this behavior is reorganized rather than simply degraded: the landscape separates into a short-time high-power branch, a narrow high-efficiency but strongly fluctuating branch, and a long-time reliability branch approaching the reset-like regime. Thus the finite-time engine cannot be characterized by a single optimal point; its useful operation depends on the intended compromise between output, efficiency, and relative power noise.

The coherence-sensitive analysis adds a further diagnostic layer. In the positive-temperature reference, the maximum-power and maximum-efficiency sectors occur in an almost decohered region of the finite-time landscape. In the population-inverted case, by contrast, the high-efficiency branch remains aligned with the dominant post-hot-bath coherence ridge. This does not imply that larger coherence is universally beneficial, nor that the engine operates with a large absolute coherence resource. Rather, it identifies where a coherence-preserving reconstruction such as DBN is operationally most relevant.

The scope of the conclusions is intentionally restricted. The inverted channel is an active reservoir-engineering primitive, and the pump or preparation cost required to maintain it has not been included. The reported efficiencies and powers are therefore gross working-medium quantities, not net resource-accounted device efficiencies. Similarly, the regularized efficiency proxy used in the scans is a normalized stochastic work output and does not replace the full stochastic efficiency ratio or its singular tail structure. These qualifications are essential for interpreting the results as design diagnostics rather than as a complete thermodynamic device claim.

From a quantum-device perspective, the main output of this work is therefore a benchmarking framework for engineered qubit thermal machines with active reservoirs. The maps identify which operating sectors are promising for large gross output, which sectors reduce relative power noise, and which sectors require coherence-preserving reconstruction to capture the relevant dynamics. Natural extensions include explicit pump or synthetic-reservoir models, resource-accounted efficiencies, sampling-error propagation for DBN tomography, and analogous finite-time operating maps in multilevel or many-body working media.

\begin{acknowledgments}
This study was financed in part by the Coordenação de Aperfeiçoamento de Pessoal de Nível Superior - Brasil (CAPES) - Finance Code 001, and by the National Natural Science Foundation of China (NSFC) under Grant No. 12174346. N.G.A. acknowledges support from FAPESP under Grant No. 2024/21707-0.
\end{acknowledgments}

\appendix

\section{DBN path construction for the five-corner Otto cycle}
\label{app:dbn_explicit}

In this appendix, we detail the explicit mathematical construction of the DBN path probabilities. We first derive the hidden corner-projector distribution $p(\alpha_1,\ldots,\alpha_5)$ directly from the underlying cycle maps, which in turn yields the inferred energy-history distribution $p_{\rm DBN}(j_1,\ldots,j_5)$. The procedure follows the operational framework of dynamic Bayesian networks: the physically measured variables correspond to projectors onto the instantaneous eigenbases of the corner density operators, while the local corner energies are subsequently inferred a posteriori via conditional probabilities evaluated by Bayes' rule. This establishes the same coherent-engine fluctuation formulation employed in Ref.~\cite{Aguilar2026}.

We consider one stationary Otto cycle described by the sequence of corner states
\begin{equation}
\rho_1 \rightarrow \rho_2 \rightarrow \rho_3 \rightarrow \rho_4 \rightarrow \rho_5,
\end{equation}
with
\begin{align}
\rho_2 &= \mathcal{M}_{1\to2}(\rho_1)
       = U_{c\to h}\rho_1 U_{c\to h}^{\dagger},
\\
\rho_3 &= \mathcal{M}_{2\to3}(\rho_2)
       = \mathcal{E}_h(\rho_2),
\\
\rho_4 &= \mathcal{M}_{3\to4}(\rho_3)
       = U_{h\to c}\rho_3 U_{h\to c}^{\dagger},
\\
\rho_5 &= \mathcal{M}_{4\to5}(\rho_4)
       = \mathcal{E}_c(\rho_4),
\end{align}

and with $\rho_1$ chosen as the fixed point of the full cycle map,
\begin{equation}
\Lambda
=
\mathcal{E}_c \circ \mathcal{U}_{h\to c} \circ \mathcal{E}_h \circ \mathcal{U}_{c\to h},
\qquad
\Lambda(\rho_1)=\rho_1.
\label{eq:app_cycle_fixed_point}
\end{equation}

At each corner $a=1,\dots,5$, we write the spectral decomposition of the corner state as
\begin{equation}
\rho_a=\sum_{\alpha_a}\lambda_{\alpha_a}^{(a)}P_{\alpha_a}^{(a)},
\label{eq:app_rho_corner_spec}
\end{equation}
and that of the corresponding corner Hamiltonian as
\begin{equation}
H_a=\sum_{j_a} e_{j_a}^{(a)} \Pi_{j_a}^{(a)}.
\label{eq:app_H_corner_spec}
\end{equation}
The DBN hidden variables are the indices $\alpha_a$ associated with the projectors $P_{\alpha_a}^{(a)}$. The energy indices $j_a$ are not measured directly; instead, they are inferred conditionally once the hidden trajectory has been specified. This is precisely the Bayesian-network viewpoint adopted in Refs.~\cite{Micadei2020,Aguilar2026}. 

The first ingredient is the probability of the initial hidden outcome,
\begin{equation}
p_1(\alpha_1)
=
\Tr\!\left[P_{\alpha_1}^{(1)}\rho_1\right]
=
\lambda_{\alpha_1}^{(1)}.
\label{eq:app_p1_alpha1}
\end{equation}
Once the outcome $\alpha_a$ at corner $a$ is given, we associate to it the normalized conditional state
\begin{equation}
\sigma_{\alpha_a}^{(a)}
=
\frac{P_{\alpha_a}^{(a)}\rho_a P_{\alpha_a}^{(a)}}
{\Tr[P_{\alpha_a}^{(a)}\rho_a]}.
\label{eq:app_sigma_alpha}
\end{equation}
For nondegenerate corner states, $\sigma_{\alpha_a}^{(a)}$ reduces to the rank-one eigenprojector itself, but the above form remains valid in the presence of degeneracies and is the one implemented numerically.

The conditional transition probability between hidden outcomes at successive corners is then obtained by propagating the conditional state through the corresponding branch map and evaluating the probability of the next projector outcome:
\begin{align} & \nonumber
T_{b|a}(\alpha_b|\alpha_a)
=
\Tr\!\left[
P_{\alpha_b}^{(b)}
\,
\mathcal{M}_{a\to b}\!\left(\sigma_{\alpha_a}^{(a)}\right)
\right],
\\
&(a,b)\in\{(1,2),(2,3),(3,4),(4,5)\}.
\label{eq:app_transition_hidden}
\end{align}
Explicitly,
\begin{align}
T_{2|1}(\alpha_2|\alpha_1)
&=
\Tr\!\left[
P_{\alpha_2}^{(2)}
U_{c\to h}\sigma_{\alpha_1}^{(1)}U_{c\to h}^{\dagger}
\right],
\\
T_{3|2}(\alpha_3|\alpha_2)
&=
\Tr\!\left[
P_{\alpha_3}^{(3)}
\mathcal{E}_h\!\left(\sigma_{\alpha_2}^{(2)}\right)
\right],
\\
T_{4|3}(\alpha_4|\alpha_3)
&=
\Tr\!\left[
P_{\alpha_4}^{(4)}
U_{h\to c}\sigma_{\alpha_3}^{(3)}U_{h\to c}^{\dagger}
\right],
\\
T_{5|4}(\alpha_5|\alpha_4)
&=
\Tr\!\left[
P_{\alpha_5}^{(5)}
\mathcal{E}_c\!\left(\sigma_{\alpha_4}^{(4)}\right)
\right].
\end{align}

The hidden five-corner trajectory probability therefore factorizes as the Markov chain
\begin{equation}
\begin{aligned}
p(\alpha_1,\alpha_2,\alpha_3,\alpha_4,\alpha_5) =
&\; p_1(\alpha_1) \,
   T_{2|1}(\alpha_2|\alpha_1) \,
   T_{3|2}(\alpha_3|\alpha_2) \\
&\times \; T_{4|3}(\alpha_4|\alpha_3) \,
   T_{5|4}(\alpha_5|\alpha_4).
\end{aligned}
\label{eq:app_hidden_path_factorization}
\end{equation}
Equation~\eqref{eq:app_hidden_path_factorization} is the explicit object denoted schematically as $p(\alpha_1,\ldots,\alpha_5)$ in the main text. In the code, these factors are implemented as the arrays $p_1$, $T_{12}$, $T_{23}$, $T_{34}$, and $T_{45}$, and their product defines the tensor of hidden-path probabilities. This makes the conditional-independence structure of the DBN completely explicit.

The second ingredient is the conditional probability of assigning a corner energy $e_{j_a}^{(a)}$ once the hidden outcome $\alpha_a$ is fixed. Using the Born rule in the conditional state $\sigma_{\alpha_a}^{(a)}$, we define
\begin{equation}
p\!\left(e_{j_a}^{(a)} \middle| \alpha_a\right)
=
\frac{
\Tr\!\left[\Pi_{j_a}^{(a)}P_{\alpha_a}^{(a)}\rho_a P_{\alpha_a}^{(a)}\right]
}{
\Tr\!\left[P_{\alpha_a}^{(a)}\rho_a\right]
}.
\label{eq:conditional_energy_given_state_1}
\end{equation}
The inferred energy-history distribution is then obtained by marginalizing over all hidden trajectories,
\begin{equation}
p_{\rm DBN}(j_1,\ldots,j_5)
=
\sum_{\alpha_1,\ldots,\alpha_5}
p(\alpha_1,\ldots,\alpha_5)
\prod_{a=1}^{5}
p\!\left(e_{j_a}^{(a)} \middle| \alpha_a\right).
\label{eq:app_pdbn_path}
\end{equation}
Equation~\eqref{eq:app_pdbn_path} is the central probabilistic object of our fluctuation analysis. It is the direct five-corner generalization of the DBN conditional-trajectory probabilities introduced for correlated heat exchange and of the Bayesian-network work distributions recently constructed for coherent Otto engines. 

Once $p_{\rm DBN}(j_1,\ldots,j_5)$ is known, all stochastic thermodynamic variables are assigned by the corner-energy differences,
\begin{align}
W_{c\to h}(\mathbf j) &= e_{j_2}^{(2)}-e_{j_1}^{(1)},
\\
Q_h(\mathbf j) &= e_{j_3}^{(3)}-e_{j_2}^{(2)},
\\
W_{h\to c}(\mathbf j) &= e_{j_4}^{(4)}-e_{j_3}^{(3)},
\\
Q_c(\mathbf j) &= e_{j_5}^{(5)}-e_{j_4}^{(4)},
\end{align}
with total stochastic work
\begin{equation}
W(\mathbf j)=W_{c\to h}(\mathbf j)+W_{h\to c}(\mathbf j).
\label{eq:app_total_work}
\end{equation}
Branchwise energetic additivity is therefore exact by construction.

A useful point to make explicit is that the cycle first law is not imposed pathwise in the present inference scheme. Indeed, by direct telescoping,
\begin{equation}
W(\mathbf j)+Q_h(\mathbf j)+Q_c(\mathbf j)
=
e_{j_5}^{(5)}-e_{j_1}^{(1)}
\equiv
\Delta E_{\rm cyc}(\mathbf j).
\label{eq:app_cycle_residual}
\end{equation}
Thus, strict pathwise closure of the cycle would require $j_5=j_1$ for every inferred history, which is not enforced by the Bayesian inference itself. What is recovered in the stationary regime is the average first law,
\begin{align}
\langle W\rangle+\langle Q_h\rangle+\langle Q_c\rangle
=&
\sum_{\mathbf j}
\Delta E_{\rm cyc}(\mathbf j)\,
p_{\rm DBN}(\mathbf j) \\
&=
\Tr[H_c(\rho_5-\rho_1)]
\simeq 0,
\label{eq:app_average_cycle_first_law}
\end{align}
up to the numerical tolerance used to determine the fixed point $\rho_1$. In other words, the DBN construction used here guarantees exact branchwise additivity and average-level thermodynamic consistency, while the cyclic closure of the inferred trajectories is only statistical. This is fully consistent with the DBN literature, where the minimally invasive character of the protocol refers to averaged measured states and averaged energetics, rather than to an exact pathwise cyclic constraint. 

Equations~\eqref{eq:app_hidden_path_factorization}--\eqref{eq:app_pdbn_path} define the complete five-corner DBN procedure implemented in our numerical code:
\begin{enumerate}[label=(\roman*)]
    \item determine the stationary corner states \(\rho_a\) from the cycle fixed-point condition,
    \item diagonalize each \(\rho_a\) and the corresponding Hamiltonian \(H_a\),
    \item construct the hidden transition kernels \(T_{2|1},\,T_{3|2},\,T_{4|3},\,T_{5|4}\) from the branch CPTP maps,
    \item form the hidden-path tensor \(p(\alpha_1,\dots,\alpha_5)\),
    \item marginalize over the hidden variables using Eq.~\eqref{eq:pdbn_path} to obtain the inferred energy-history distribution \(p_{\mathrm{DBN}}(j_1,\dots,j_5)\).
\end{enumerate}
Because the thermalization strokes are described by the time-independent Lindblad master equation~\eqref{eq:ME}, the corresponding maps are formally given by \(\mathcal{E}_\alpha = e^{\mathcal{L}_\alpha \tau_\alpha}\), where \(\mathcal{L}_\alpha\) is the standard Liouvillian superoperator.
This amplitude-damping-type generator ensures that \(\mathcal{E}_{h,c}\) are completely positive and trace-preserving (CPTP), modeling a physically and thermodynamically consistent dissipative process rather than a phenomenological interpolation.

\section{Numerical implementation}
\label{app:numerics}

Numerical simulations were performed using the QuTiP library \cite{Johansson2012,Johansson2013} in units of $\hbar=1$. The working-medium Hamiltonians are $H_c=-(\omega_c/2)\sigma_x$ and $H_h=-(\omega_h/2)\sigma_y$. All physical parameters, parameter grids, and ODE solver tolerances are summarized in Table~\ref{tab:numerics}.

For the finite-time landscapes, we imposed symmetric hot and cold isochore durations,
\begin{equation}
	\tau_h=\tau_c\equiv\tau_{\rm iso}.
\end{equation}
The stationary limit cycle was evaluated iteratively until the fixed-point condition $\|\rho_{n+1}-\rho_n\|_1<10^{-12}$ was satisfied. The ideal-reset benchmark was recovered by setting $\tau_{\rm iso}=5\times10^4$, ensuring trace distances to the exact stationary states below $10^{-12}$. For the DBN--TPM comparison in Appendix~\ref{app:dbn_tpm_benchmark}, we fixed $\tau_{\rm iso}=100$ and performed a one-dimensional scan over $\tau_{\rm drive}$.

We verified that refining the time discretization, increasing the grid densities, or imposing stricter solver tolerances does not alter the qualitative structure of the reported operating sectors.

\begin{table}[t]
	\caption{Physical parameters and numerical grid settings used in the simulations.}
	\label{tab:numerics}
	\centering
	\begin{ruledtabular}
		\begin{tabular}{lc}
			\textbf{Physical parameters} & \\
			\hline
			Cold gap, $\omega_c$ & $0.0157$ \\
			Hot gap, $\omega_h$ & $0.0314$ \\
			Reference rate, $\gamma_{\rm ref}$ & $10^{-3}$ \\
			Populations, inverted case, $(p_e^c,p_e^h)$ & $(0.40,0.80)$ \\
			Populations, positive-temperature case, $(p_e^c,p_e^h)$ & $(0.25,0.35)$ \\
			\vspace{-2mm}\\
			\textbf{Numerical settings} & \\
			\hline
			$\tau_{\rm drive}$ grid, logarithmic points & $80\in[20,5000]$ \\
			$\tau_{\rm iso}$ grid, logarithmic points & $100\in[5,5\times10^4]$ \\
			Steps, $(n_{\rm steps}^{\rm drive},n_{\rm steps}^{\rm iso})$ & $(500,400)$ \\
			Solver tolerances, $(\mathtt{atol},\mathtt{rtol})$ & $10^{-10}$ \\
		\end{tabular}
	\end{ruledtabular}
\end{table}

For the parameter set used in the simulations, the intrinsic timescales of the model are set by the oscillation periods
\begin{equation}
	T_c=\frac{2\pi}{\omega_c}\approx 400,
	\qquad
	T_h=\frac{2\pi}{\omega_h}\approx 200,
\end{equation}
and by the relaxation time $\tau_{\rm rel}\sim\gamma_{\rm ref}^{-1}\approx10^3$. Within the present effective model, $\tau_{\rm rel}$ may be identified with the order of the qubit $T_1$. In these units, the full-thermalization benchmark uses $\tau_{\rm iso}=5\times10^4$, corresponding to about $50\,\tau_{\rm rel}$, and therefore represents a deliberately conservative long-time limit. The finite-time scans cover
\begin{equation}
	\tau_{\rm drive}\in[20,5000]\approx[0.1T_h,25T_h],
\end{equation}
and
\begin{equation}
	\tau_{\rm iso}\in[5,5\times10^4]\approx[5\times10^{-3}\tau_{\rm rel},50\tau_{\rm rel}],
\end{equation}
thereby spanning the crossover from strongly nonadiabatic, weakly thermalized cycles to near-full thermalization.

\section{Positive-temperature finite-time reference landscape}
\label{app:positive_reference}

\begin{figure*}[t!]
	\centering
	\includegraphics[width=1.05\textwidth]{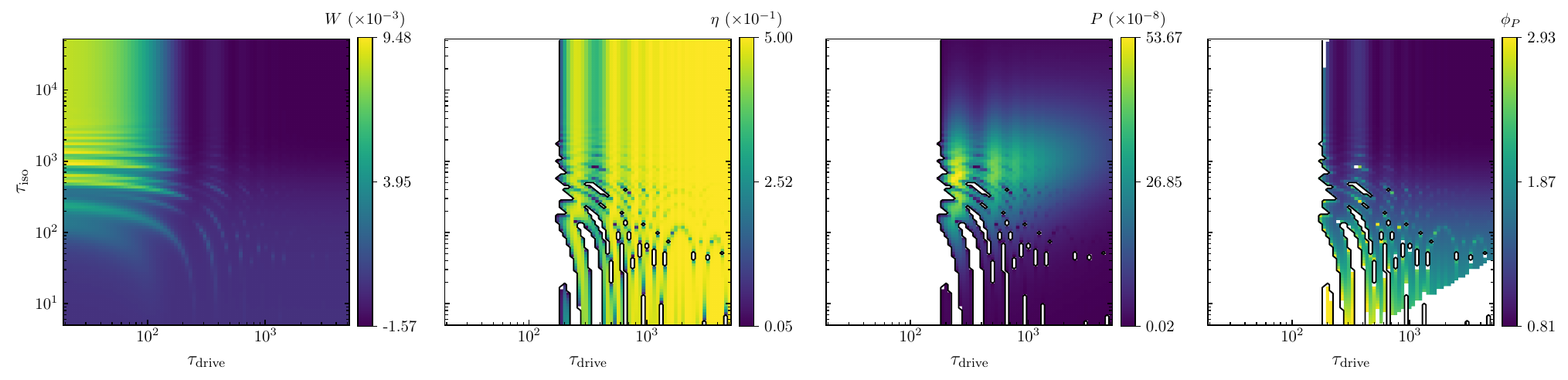}
	\caption{
		Finite-time operating landscape for the positive-temperature reference case in the $(\tau_{\rm drive},\tau_{\rm iso})$ plane, showing the extracted work $\langle W\rangle$, efficiency $\eta_{\rm reg}$, output power $P$, and relative power fluctuation $\phi_P=\sigma(P)/|\langle P\rangle|$. This figure is included as an appendix reference baseline for comparison with the population-inverted operating landscape discussed in the main text and with the coherence-sensitive analysis. The white region corresponds to points outside the engine regime.
	}
	\label{fig:positive_finite_reference}
\end{figure*}

For reference, Fig.~\ref{fig:positive_finite_reference} shows the finite-time operating landscape of the positive-temperature case used as the baseline in the main-text analysis. The figure displays the extracted work, efficiency, output power, and relative power fluctuation in the $(\tau_{\rm drive},\tau_{\rm iso})$ plane. Its purpose is to provide the direct positive-temperature counterpart of the population-inverted finite-time landscape discussed in Sec.~\ref{sec:finite_time_landscapes}. In particular, it makes clear that the broad structured operating window found in the inverted reduced model has no direct analogue in the positive-temperature reference case.

\section{Coherence comparison between positive- and negative-temperature regimes}
\label{app:coherence}

\begin{figure*}[t!]
	\centering
	\includegraphics[width=0.95\textwidth]{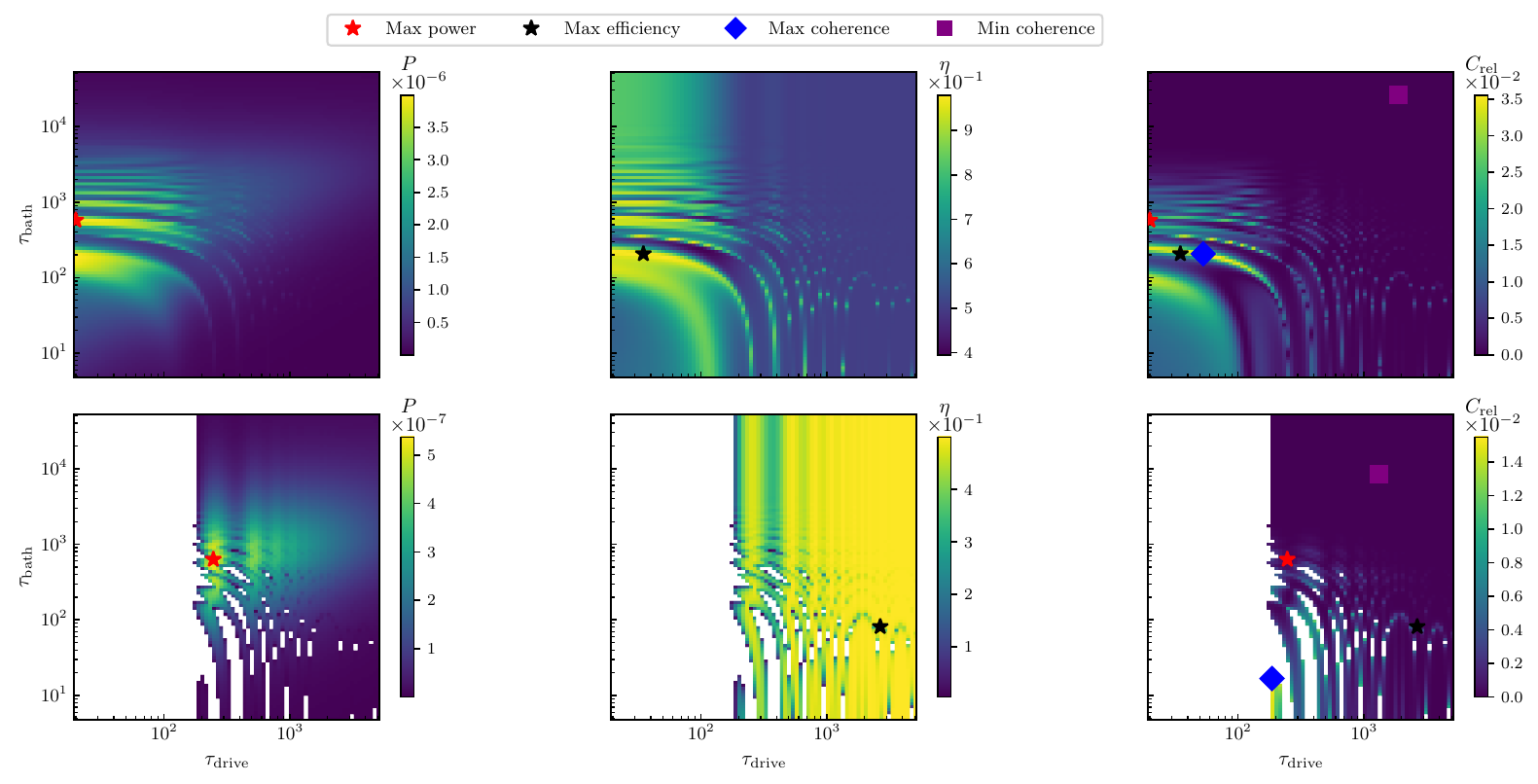}
	\caption{
		Finite-time thermodynamic and coherence landscapes in the $(\tau_{\rm drive},\tau_{\rm iso})$ plane. The top row corresponds to the effective negative-temperature regime, while the bottom row shows the positive-temperature reference case. The columns display the output power $P$, the efficiency $\eta_{\rm reg}$, and the relative-entropy coherence after the hot bath, $C_{\rm rel}(\rho_3;H_h)$. White regions indicate points outside the engine regime. In each row, the markers indicate the maximum-power point, the maximum-efficiency point, the maximum-coherence point, and the minimum-coherence point. The figure makes explicit that the positive-temperature optimal sectors lie in an almost fully decohered region, whereas in the population-inverted regime the maximum-efficiency branch remains aligned with the dominant coherence ridge.
	}
	\label{fig:coherence_landscape_comparison}
\end{figure*}

To make the coherence-sensitive structure of the finite-time regime explicit, Fig.~\ref{fig:coherence_landscape_comparison} compares the efficiency $\eta_{\rm reg}$, the output power $P$, and the post-hot-bath coherence $C_{\rm rel}(\rho_3;H_h)$ in the $(\tau_{\rm drive},\tau_{\rm iso})$ plane for the population-inverted engine and for the positive-temperature reference. This comparison clarifies two points that are only indirectly visible from Table~\ref{tab:coherence_summary}. First, the absolute coherence scale remains modest in both cases when compared with the qubit upper bound $\ln 2$. Second, the role of coherence is strongly regime dependent: in the positive-temperature reference, the optimal operating sectors lie in an almost fully decohered region, whereas in the inverted case the maximum-efficiency branch remains aligned with the dominant coherence ridge of the landscape.

The point is therefore not that the reduced population-inverted engine operates with an exceptionally large absolute coherence resource. Rather, its useful finite-time high-efficiency sector remains structurally connected to the leading surviving coherence after the hot isochore. This supports the interpretation used in the main text: the coherence map is an operational diagnostic of where a coherence-preserving reconstruction is expected to matter most.

\section{DBN--TPM benchmark in the finite-time population-inverted regime}
\label{app:dbn_tpm_benchmark}

Since the DBN framework is used in the main text as the reconstruction protocol for the coherent, unmeasured cycle, it is useful to compare it explicitly with the standard TPM protocol in a representative finite-time setting. To this end, we consider a fixed population-inverted working point and perform a one-dimensional scan in the driving time $\tau_{\rm drive}$, keeping $\tau_h=\tau_c\equiv\tau_{\rm iso}$ fixed. For each value of $\tau_{\rm drive}$, we compute: (i) the exact coherent-cycle averages obtained from the unmeasured dynamics, (ii) the DBN reconstruction, and (iii) the TPM benchmark built from explicit projective energy measurements at the stroke boundaries.

\begin{figure*}[t!]
	\centering
	\includegraphics[width=0.85\textwidth]{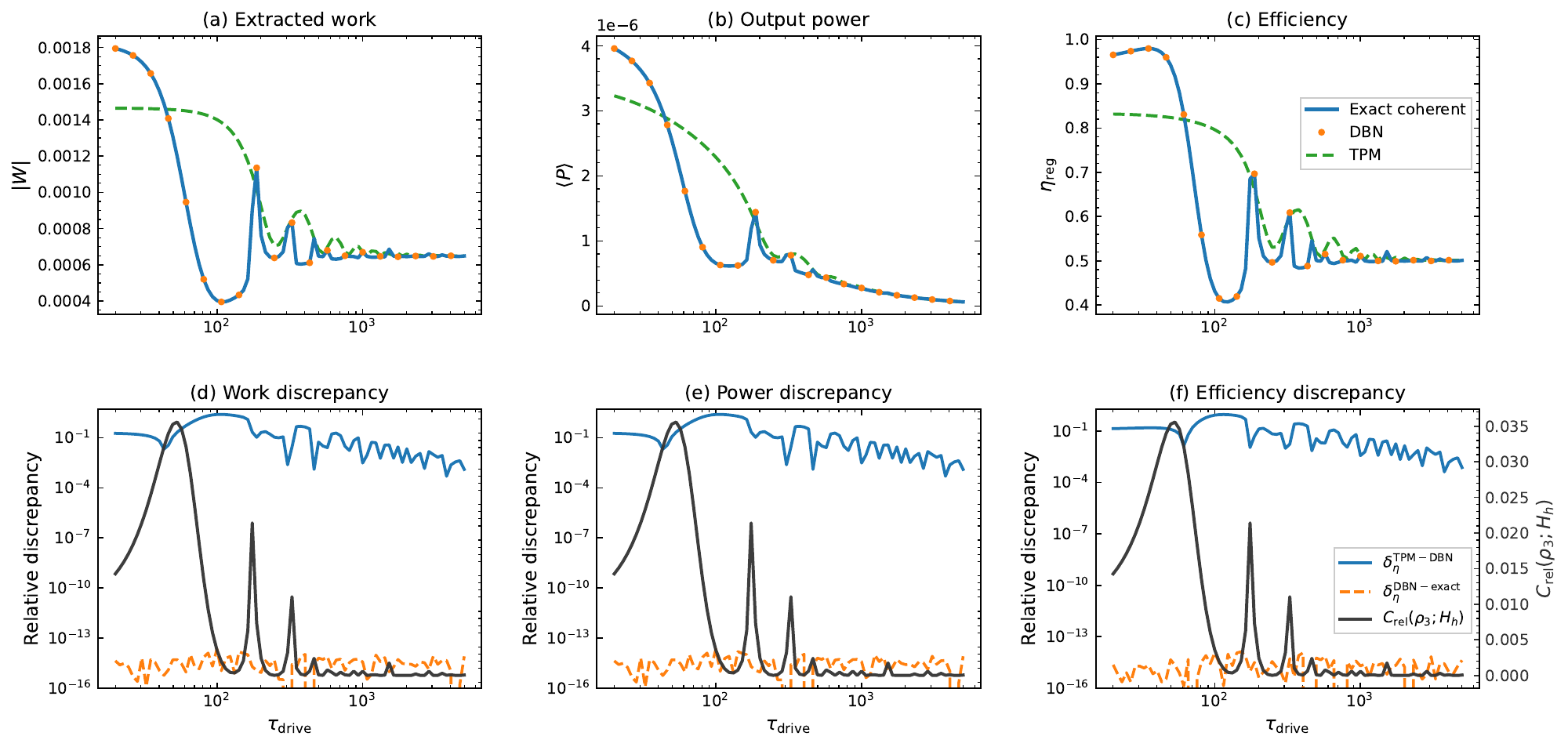}
	\caption{
		Benchmark of the dynamic Bayesian network (DBN) reconstruction against the standard two-point measurement (TPM) protocol in the finite-time population-inverted regime. Panels (a)--(c) show the mean extracted work $|\langle W\rangle|$, output power $\langle P\rangle$, and efficiency $\eta_{\rm reg}$ as functions of the driving time $\tau_{\rm drive}$ at fixed isochore duration $\tau_{\rm iso}$. The coherent-cycle averages are reproduced by the DBN reconstruction to numerical precision, while the TPM predictions deviate significantly in the short-time regime. Insets provide a zoomed-in view of the crossover region. Panels (d)--(f) show the relative discrepancies for the corresponding thermodynamic observables. The DBN reconstruction error $\delta^{\rm DBN-exact}$ remains at machine precision across the scan, confirming the average-preserving property of the reconstruction for this model. The TPM deviation $\delta^{\rm TPM-DBN}$ is pronounced at short driving times and correlates with the relative-entropy coherence $C_{\rm rel}(\rho_3;H_h)$ surviving after the hot bath. Both descriptions converge in the long-time limit, where nonadiabatic excitations and residual coherences are strongly suppressed.
	}
	\label{fig:dbn_tpm_benchmark}
\end{figure*}

Figure~\ref{fig:dbn_tpm_benchmark} shows the resulting comparison. Two features are immediately clear. First, the DBN averages reproduce the exact coherent-cycle averages to numerical precision throughout the full scan. Second, the TPM benchmark departs significantly from DBN in the short-time regime, precisely where the coherence after the hot isochore is largest. As $\tau_{\rm drive}$ increases and the cycle becomes progressively less coherence active, the TPM and DBN results converge.

This comparison clarifies the role of the DBN construction in the present problem. The difference between DBN and TPM is not a numerical artifact, but the direct consequence of the projective dephasing built into TPM. In the long-time regime, where coherence is weak, that dephasing becomes essentially irrelevant and both descriptions agree. In the short-time regime, however, where coherence remains thermodynamically active, TPM no longer reproduces the same cycle averages as the coherence-preserving dynamics.

This benchmark is deliberately limited to average extracted work, power, and efficiency. A complete distribution-level comparison of DBN and TPM work--heat statistics would require additional numerical data, including the joint distributions $P(W,Q_h)$ and the corresponding stochastic-efficiency ratio $-W/Q_h$. The present result therefore supports the narrower interpretation used in the main text: DBN is the appropriate reconstruction for the untouched coherent cycle, whereas TPM describes a different, explicitly measured and dephased engine.

A distribution-level DBN--TPM comparison, for example using distances between $P(W)$ or joint distributions $P(W,Q_h)$ at selected operating points, would be a natural extension of the present average-level benchmark.

\section{Bound diagnostics}
\label{app:bounds}

The information-theoretic diagnostics discussed in the DBN literature are used here only as consistency checks, not as an independent central result of the paper. Since the main text focuses on the population-inverted engine and its finite-time coherence-sensitive structure, we restrict the bound analysis to that regime.

For each thermalization stroke, we evaluate the change in the sum of an energy-basis coherence contribution $C$ and a commuting athermality contribution $D$. Here $C$ is the same relative-entropy coherence introduced in Sec.~\ref{sec:coherence_finite},
\begin{equation}
	C(\rho;H)
	\equiv
	S(\Delta_H[\rho])-S(\rho),
	\label{eq:C_def_app}
\end{equation}
where $\Delta_H[\rho]$ denotes complete dephasing in the eigenbasis of $H$. The commuting athermality contribution is
\begin{equation}
	D(\rho;H,\beta)
	=
	S\!\left(\Delta_H[\rho]\middle\|\rho_\beta\right),
	\qquad
	\rho_\beta=
	\frac{e^{-\beta H}}{\Tr(e^{-\beta H})},
	\label{eq:D_def_app}
\end{equation}
with $S(\rho\|\sigma)=\Tr[\rho(\ln\rho-\ln\sigma)]$. Thus $C$ quantifies the coherent contribution in the energy basis, while $D$ quantifies the residual population athermality once coherence has been removed.

Using the corner states of the Otto cycle, we define
\begin{align}
	\Delta(C+D)_{\rm hot}
	=&
	\left[
	C(\rho_3;H_h)+D(\rho_3;H_h,\beta_h)
	\right]
	\nonumber\\
	&-
	\left[
	C(\rho_2;H_h)+D(\rho_2;H_h,\beta_h)
	\right],
	\label{eq:DeltaCD_hot}
\end{align}
and
\begin{align}
	\Delta(C+D)_{\rm cold}
	=&
	\left[
	C(\rho_5;H_c)+D(\rho_5;H_c,\beta_c)
	\right]
	\nonumber\\
	&-
	\left[
	C(\rho_4;H_c)+D(\rho_4;H_c,\beta_c)
	\right].
	\label{eq:DeltaCD_cold}
\end{align}
In the present constant-Hamiltonian thermalization strokes, the relevant monotonicity condition reduces to
\begin{equation}
	\Delta(C+D)\le 0.
	\label{eq:monotonicity_cond}
\end{equation}

\begin{figure}[t!]
	\centering
	\includegraphics[width=0.50\textwidth]{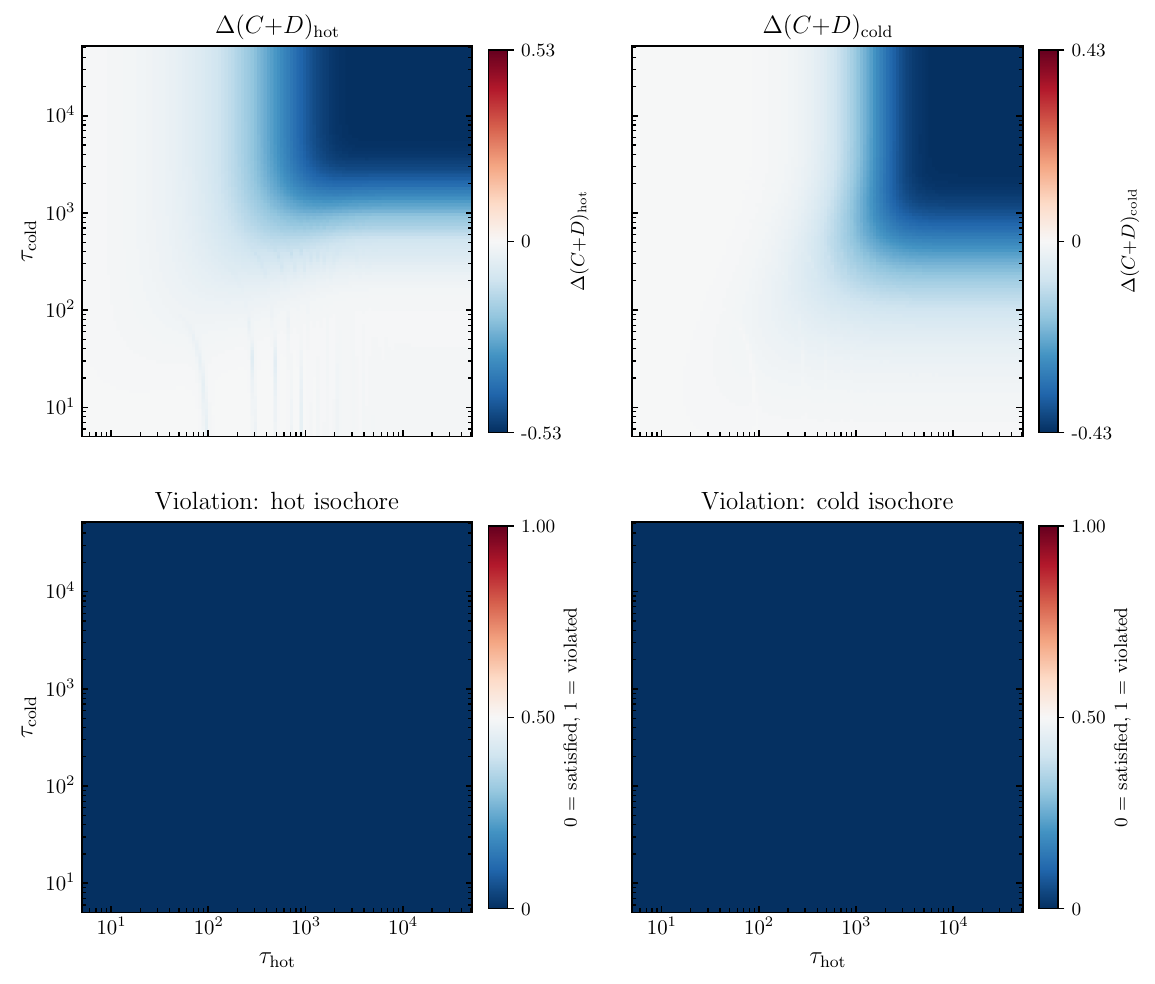}
	\caption{
		Finite-time bound diagnostics for the population-inverted engine. Left panels: $\Delta(C+D)_{\rm hot}$ and $\Delta(C+D)_{\rm cold}$ as defined in Eqs.~\eqref{eq:DeltaCD_hot} and \eqref{eq:DeltaCD_cold}. Right panels: corresponding binary violation maps, where $0$ indicates that the bound $\Delta(C+D)\le0$ is satisfied and $1$ indicates violation. The black contour marks the engine boundary; white areas lie outside the engine region. Throughout the explored parameter space, no violation occurs, i.e., the inequalities $\Delta(C+D)_{\rm hot}\le0$ and $\Delta(C+D)_{\rm cold}\le0$ hold everywhere.
	}
	\label{fig:bounds_negative}
\end{figure}

We compute these quantities over the population-inverted finite-time parameter space and construct the corresponding binary violation maps. The numerical result is straightforward: throughout the engine region explored in the inverted scan, both inequalities remain satisfied,
\begin{equation}
	\Delta(C+D)_{\rm hot}\le0,
	\qquad
	\Delta(C+D)_{\rm cold}\le0.
	\label{eq:inequalities_satisfied}
\end{equation}
No point in the physically relevant inverted-engine sector exhibits a breakdown of the expected information-theoretic monotonicity of the hot- and cold-bath strokes. The values of $\Delta(C+D)_{\rm hot}$ and $\Delta(C+D)_{\rm cold}$ typically become more negative in the interior of the allowed operating region and approach zero only near the boundaries, without changing sign.

This observation should be interpreted as a robustness check on the finite-time reconstruction rather than as a new thermodynamic claim. The population-inverted finite-time regime is precisely the part of the paper where coherence remains visibly active and the performance landscape is most strongly structured. The absence of violations therefore confirms that the coherence-active operating sectors discussed in the main text remain consistent with the expected information-theoretic monotonicity of the thermalization dynamics.

Although thermodynamic-uncertainty-relation-like bounds are often used to analyze power and stochastic-efficiency fluctuations, we do not adopt them here as the main consistency test. Their standard derivations do not explicitly account for coherence-preserving fluctuation reconstructions of the DBN type, and the regularized efficiency proxy used in our scans is not the literal stochastic ratio. In the present setting, a naive comparison with a standard TUR could therefore produce an apparent bound violation whose interpretation would be ambiguous. We instead use the $C+D$ monotonicity diagnostics established in the DBN literature, which are tailored to coherence-sensitive finite-time dynamics.

\bibliography{refs_aqt}

\end{document}